\documentclass[namedreferences]{solarphysics}

\usepackage[hyperref,optionalrh]{spr-sola-addons} % For Solar Physics 
\usepackage{graphicx}        % For eps, pdf figures
\usepackage[usenames,dvipsnames]{color} % For color text: \color command
\usepackage{breakurl}        % For breaking URLs easily trough lines
\usepackage{relsize}

\newcommand{\BE}{\begin{equation}}
\newcommand{\EE}{\end{equation}}
\newcommand{\BA}{\begin{align}}
\newcommand{\EA}{\end{align}}
 \newcommand{\fig}[1]{Figure~\ref{fig_#1}}
 \newcommand{\figS}[1]{Figures~\ref{fig_#1}}
 
 \newcommand{\figs}[2]{Figures~\ref{fig_#1} and \ref{fig_#2}}
 
 \newcommand{\sect}[1]{Section~\ref{sect_#1}}
 
 \newcommand{\eq}[1]{Equation~\ref{eq_#1}}
 \newcommand{\eqs}[2]{Equations~\ref{eq_#1} and \ref{eq_#2}}
 \newcommand{\eqss}[2]{Equations~\ref{eq_#1} - \ref{eq_#2}}

\newcommand{\eg}{{\it e.g.},}

\newcommand{\ie}{{\it i.e.},}
\newcommand{\insitu}{{\it in situ}}
\newcommand{\versus}{{\it versus}}
\newcommand{\Int}[2]{\ensuremath{\mathchoice%
        {\displaystyle\int_{#1}^{#2}}
        {\displaystyle\int_{#1}^{#2}}
        {\int_{#1}^{#2}}
        {\int_{#1}^{#2}}
        }}

\newcommand{\degree}{\ensuremath{^\circ}}

\newcommand{\rmd}{{\rm d }}
\newcommand{\uvec}[1]{\hat{ \textit{\textbf{#1}} }}
\renewcommand{\vec}[1]{ \textit{\textbf{#1}} }

\newcommand{\asf}{\rm asf}
\newcommand{\bA}{\beta_{\rm A}}
\newcommand{\BL}{\vec{B}_{\rm L}}
\newcommand{\Bo}{B_{0}}
\newcommand{\Bobs}{\vec{B}_{\rm obs}}

\newcommand{\ba}{b_{a}}
\newcommand{\bainf}{b_{a,\rm  inf}}
\newcommand{\basup}{b_{a,\rm  sup}}
\newcommand{\bz}{b_{z}}
\newcommand{\bzinf}{b_{z,\rm  inf}}
\newcommand{\bzsup}{b_{z,\rm  sup}}
\newcommand{\Ba}{B_{a}}

\newcommand{\Bz}{B_{z}}

\newcommand{\BzFR}{B_{ z,\rm FR}}

\newcommand{\chiB}{\mathlarger{\mathlarger{\chi}}_{\rm B}}  
\newcommand{\chiR}{\mathlarger{\mathlarger{\chi}}_{\rm R}}

\newcommand{\dt}{\Delta t}
\newcommand{\ea}{\uvec{e}_{a}}
\newcommand{\ez}{\uvec{e}_{z}}
\newcommand{\fa}{f_{a}}
\newcommand{\Fa}{F_{a}}
\newcommand{\fainf}{f_{a,\rm inf}}
\newcommand{\fasup}{f_{a,\rm sup}}
\newcommand{\fz}{f_{z}}
\newcommand{\Fz}{F_{z}}
\newcommand{\fzinf}{f_{z,\rm inf}}
\newcommand{\fzsup}{f_{z,\rm sup}}
 
\newcommand{\hinf}{h_{\rm inf}} 
 
\newcommand{\hsup}{h_{\rm sup}}

\newcommand{\lA}{\lambda}
 
\newcommand{\pA}{\phi} %_{\rm A}}
\newcommand{\pinf}{p_{\rm inf}} 
\newcommand{\psup}{p_{\rm sup}} 
\newcommand{\Rin}{R_{\rm in}} 
\newcommand{\rinf}{r_{\rm inf}} 
\newcommand{\Rinf}{R_{\rm inf}} 
\newcommand{\Rout}{R_{\rm out}} 
\newcommand{\Ro}{R_{0}}
\newcommand{\Rsup}{R_{\rm sup}} 
\newcommand{\rsup}{r_{\rm sup}} 
\newcommand{\tA}{\theta} %_{\rm A}} 

\renewcommand{\to}{t_{0}}
\newcommand{\Xin}{X_{\rm in}} 
\newcommand{\Xout}{X_{\rm out}} 

\newcommand{\Vmean}{\left< V \right>}
\newcommand{\Wind}{{\it Wind}}

\begin{document}

\begin{article}
\begin{opening}

\title{Re-analysis of Lepping's Fitting Method for Magnetic Clouds: 
Lundquist Fit Reloaded}

   % authors 
\author[addressref=aff1, corref, email={Pascal.Demoulin@obspm.fr}]
       {\inits{P.}\fnm{Pascal}~\lnm{D\'emoulin} \orcid{0000-0001-8215-6532}}
\author[addressref={aff2,aff3}, email={dasso@df.uba.ar} ]
       {\inits{S.}\fnm{Sergio}~\lnm{Dasso}      \orcid{0000-0002-7680-4721}}
\author[addressref=aff4, email={mjanvier@ias.u-psud.fr}]
       {\inits{M.}\fnm{Miho}~\lnm{Janvier}      \orcid{0000-0002-6203-5239}}
\author[addressref={aff3} ]
       {\inits{V.}\fnm{Vanina}~\lnm{Lanabere} }   
   % addresses 
\address[id=aff1]{LESIA, Observatoire de Paris, Universit\'e PSL, CNRS, 
                  Sorbonne Universit\'e, Univ. Paris Diderot, Sorbonne Paris Cite,
                  5 place Jules Janssen, 92195 Meudon, France}
\address[id=aff2]{CONICET, Universidad de Buenos Aires, Instituto de Astronom\'\i a 
                  y F\'\i sica del Espacio, CC. 67, Suc. 28, 1428 Buenos Aires,Argentina} 
\address[id=aff3]{Universidad de Buenos Aires, Facultad de Ciencias Exactas y Naturales,
                  Departamento de Ciencias de la Atm\'osfera  y los Oc\'eanos and
                  Departamento  de F\'\i sica, 1428 Buenos Aires, Argentina}
\address[id=aff4]{Institut d'Astrophysique Spatiale, UMR8617, Univ. Paris-Sud-CNRS, 
                  Universit\'e Paris-Saclay, B\^atiment 121, 91405 Orsay Cedex, France} 

\runningauthor{P. D\'emoulin \textit{et al.}}
\runningtitle{Lundquist Fit Reloaded: Application to Magnetic Clouds}

\begin{abstract}
  % generalities
Magnetic clouds (MCs) are a subset of ejecta, launched from the Sun as coronal mass ejections.  The coherent rotation of the magnetic field vector observed in MCs leads to envision MCs as formed by flux ropes (FRs).  
  % Lepping fit revisited
Among all the methods used to analyze MCs, Lepping's method (Lepping, Jones, and Burlaga, 1990, \textit{J. Geophys. Res.} \textbf{95}, 11957) is the broadest used.  While this fitting method does not require the axial field component to vanish at the MC boundaries, this idea is largely spread in publications.
Then, we revisit Lepping's method to emphasize its hypothesis and the meaning of its output parameters.  As originally defined, these parameters imply a fitted FR which could be smaller or larger than the studied MC.  We rather provide a re-interpretation of Lepping's results with a fitted model limited to the observed MC interval.  
   % main results
We find that, typically the crossed FRs are asymmetric with a larger side both in size and magnetic flux before or after the FR axis.
At the boundary of the largest side we find an axial magnetic field component distributed around zero which we justify by the physics of solar eruptions.  In contrast, at the boundary of the smaller side the axial field distribution is shifted to positive values, as expected with erosion acting during the interplanetary travel. 
  % Consequences
This new analysis of Lepping's results have several implications.
   First, global quantities, such as magnetic fluxes and helicity, need to be revised depending on the aim (estimating global properties of FRs just after the solar launch or at 1 au).
   Second, the deduced twist profiles in MCs range quasi-continuously from nearly uniform, to increasing away from the FR axis, up to a reversal near the MC boundaries.  There is no trace of outsider cases, but a continuum of cases.
   Finally, the impact parameter of the remaining FR crossed at 1 au is revised. Its distribution is compatible with weakly flatten FR cross-sections.   
\end{abstract}


\end{opening}
%-------------------------------------------------

%%%%%%%%%%%%%%%%%%%%%%%%%%%%%%%%%%%%%%%%%%%%%%%%%%%%%%%%%%%%%%%%%%%%%%%%%%%%%%%%%%
\section{Introduction}  %%%%%%%%%%%%%%%%%%%%%%%%%%%%%%%%%%%%%%%%%%%%%%%%%%%%%%%%%%
     \label{sect_Intro} 

  %%{\S}{\bf --- Intro for MCs} \\
 Magnetic clouds (MCs) are a part of ejecta launched from the Sun.  Many studies are focussed on MCs because of their highly organized magnetic field. 
The \insitu\ observations at the Earth orbit show a large and coherent rotation of the magnetic field vector on a time scale of the order of one day \citep[\eg ][]{Burlaga81,Burlaga82b,Gosling90,Lepping90}.  These observations are typically interpreted as the presence of a twisted magnetic flux tube, or flux rope (FR), crossed by the spacecraft.
MCs also have typically a lower proton temperature compared to the solar wind with the same speed, and a low plasma beta. As the magnetic forces are expected to be dominant, MCs are described with a force-free configuration in a progressive expansion as they move away from the Sun \citep[\eg ][]{Shimazu02,Demoulin09,Vandas15b}.  

  %{\S}{\bf --- From \insitu\ data to FR properties} \\
While \insitu\ observations provide detailed plasma and magnetic field measurements along the spacecraft trajectory, modeling is needed to estimate more global physical parameters, such as magnetic flux and helicity.  In particular, such approach is needed to relate the local \insitu\ observations to the physics of the coronal source region \citep[\eg ][]{Dasso05, Qiu07, Nakwacki11, Rodriguez08, Hu14}.  A classical approach is to fit a FR model to the observed magnetic field.  This allows finding approximately the local orientation of the FR axis; then, the data are rotated in the FR frame defined by the FR axis direction and the spacecraft trajectory. 
    
  %{\S}{\bf --- Classical Lundquist model} \\
Several FR models have been proposed to describe MCs.  
The simplest ones assume local cylindrical symmetric magneto-static FR solutions.
Among the ones with the lowest number of free parameters, the Lundquist's model assumes a linear force-free field (FFF) with a simple profile \citep[described with one harmonic of the linear FFF,][]{Lundquist50,Goldstein83}.  This field is at the base of Lepping's method for MC data fitting \citep{Lepping90}.
Despite its simplicity, this model describes well the magnetic field of a large fraction of MCs, at least when the unity field vectors $\vec{B}/B$ are compared \citep[\eg ][]{Lepping90, Burlaga95, Burlaga98, Lynch03, Dasso05, Lynch05, Wang15}. 
   
  %{\S}{\bf --- GH model } \\
Another simple model is a non-linear FFF with a uniform twist, known as the Gold and Hoyle's solution \citep{Gold60}.  It has a cylindrical symmetry and the same number of free parameters as the Lundquist's model.  This model was applied successfully to case studies \citep[\eg ][]{Farrugia99, Dasso03, Dasso06} and more recently to larger sets of MCs \citep{Hu15, Wang16}. 

  %{\S}{\bf --- Other models + specific interest of Lundquist model} \\
Other models, with more free parameters, have been proposed: for example, with an elliptical cross-section shape and/or non-force free fields \citep[\eg ][]{Vandas03,Hidalgo02b,Nieves18b}.   Still, the Lundquist's model with Lepping's fitting procedure remains a reference in the domain, in particular since it was applied systematically to MCs observed at 1 au by \Wind\ spacecraft, demonstrating its ability to fit a large variety of data \citep[\eg ][]{Lepping03a, Lepping06, Lepping07, Lepping10, Lepping18a}.  Several authors developed their own version based on the pioneering work's of Lepping \citep[\eg ][]{Lynch05, Leitner07, Marubashi15a, Good19, Nishimura19}.  Furthermore, it became a source of comparison for authors developing other approaches \citep[\eg ][]{Nieves05,Wang15,Nieves18b}. 

  %{\S}{\bf --- Application Lundquist model to CMEs} \\
The Lundquist's profile was also used in combination with a model having a density shell with a torus shape and fitted visually to white-light observations of a coronal mass ejection \citep{Thernisien06,Wood17}. This defines a FR model in the solar corona.  The axial magnetic field component is supposed to vanish at the FR boundary (the density shell) of the torus model and the azimuthal flux is supposed to be the reconnected flux deduced from the photospheric magnetogram and the flare loop extension \citep{Pal17,Gopalswamy17,Gopalswamy18}. 

  %{\S}{\bf --- Defining MC boundaries} \\
With \insitu\ data, MC boundaries could be difficult to define, especially the rear boundary.  Different authors are frequently considering different boundaries, typically with differences of few hours, because they consider different criteria applied on different \insitu\ parameters \citep[\eg ][]{Zurbuchen06, Dasso06, Al-Haddad13}. Since this is a research subject by itself, in present study we keep the boundaries as defined in the Lepping's table, and it is not our aim to make comparison with other
authors in this study. 
 
  %{\S}{\bf --- How the Lundquist model is fitted?} \\
In Lepping's papers, it reads that a constraint for a vanishing axial field component is imposed at the FR boundary: $\BzFR=0$ in the FR frame (with axis attached to the FR, \eg\ one axis is along and the others orthogonal to the FR axis).  While $\BzFR$ can be found small at the boundaries of some MCs, this is not true in general and the papers do not provide further justification for this imposed constraint.  In order to check the validity of this hypothesis both the data and the fitted model could be plotted in the FR frame.  However, this is typically not shown in the papers.  Then, whether Lepping's method really imposes $\BzFR =0$ at the boundaries still needs to be checked.

  %{\S}{\bf --- Doubts on imposing $\BzFR=0$} \\
 Further doubts arise as to whether this constraint, $\BzFR=0$ at the MC boundaries, is needed since some MCs show a reversal of $\BzFR$ significantly inside the boundaries \citep{Vandas01, Lepping03a, Lepping06}.  Furthermore, previous studies showed that a dominant fraction of MCs are eroded, at least on one side (front and/or rear), during the travel from the Sun to 1 au \citep[][]{Dasso06,Dasso07,Ruffenach15}.  With such erosion by magnetic reconnection with the solar wind field, how could a small axial component could remain at the eroded FR boundary?     

  %{\S}{\bf --- Implication of setting $\BzFR =0$} \\
Forcing $\BzFR =0$ at the MC boundaries could have an important effect on the deduced axis orientation for MCs where this condition is not satisfied.  This was shown by \citet{Nishimura19} who fit a large set of MCs with the Lundquist model.  They found that the deduced axis direction could be significantly different for 30\% of the MCs when the condition $\BzFR =0$ is imposed or not at these boundaries for the fitted model. 
These differences propagate in all the other parameters which depend on the FR axis orientation.   Then, when different authors compare their results on the same events, it is important to precise not only the time location of the selected MC boundaries but also if $\BzFR =0$ is imposed or not at these boundaries.   
  
  %{\S}{\bf --- Roadmap of the paper} \\
  In \sect{Fit}, we first review Lepping's fitting procedure of \insitu\ magnetic field within MCs.  We show that, while the fitting method does not require the axial magnetic field component to vanish at the MC boundaries, the selected output parameters have introduced this misleading assumption in subsequent published papers.  We define other output parameters which better represent the crossed MC. This provides deeper informations on the physics involved in the launch and transport of FRs. 
  Next, we explore in \sect{Implications} the consequences of considering only the fitted model within the observed MC boundaries.  This affects significantly previous results on magnetic flux, twist, and helicity as well as the impact parameter distribution.  Finally, our conclusions, derived from re-analysing Lepping's results, are given in \sect{Conclusion}.

%%%%%%%%%%%%%%%%%%%%%%%%%%%%%%%%%%%%%%%%%%%%%%%%%%%%%%%%%%%%%%%%%%%%%%%%%%%%%%%%%%
\section{Lepping's Fitting Method} %%%%%%%%%%%%%%%%%%%%%%%%%%%%%%%%%%%%%%%%%%%%%%%
  \label{sect_Fit}

\subsection{Interpretation of the Output Parameters} %%%%%%%%%%%%%%%%%%%%%%%%%%%%%%%
  \label{sect_Fit_Interpretation}

    %{\S}{\bf --- Lundquist's model} \\
\citet{Lepping90} and \citet{Lepping06} developed a method and a numerical tool to perform a least square fit of the magnetic field vector within MCs in order to find the FR properties such as its orientation and its central magnetic strength.  They applied it to the observations of MCs made by the \Wind\ spacecraft.
The magnetic field model, $\BL$, is the linear FFF with a cylindrical symmetry, so with a circular section and a straight axis \citep{Lundquist50}:    
  \begin{equation}  \label{eq_Lundquist}
  \BL (\rho ) = \Ba \ea +\Bz \ez 
              = \Bo \, [H \, J_1(\alpha \rho ) \, \ea + J_0(\alpha \rho ) \, \ez ] \,.
  \end{equation}
$\rho$ is the distance to the FR axis.
$\ea$ ($\ez$) is the azimuthal (axial) unit vectors in cylindrical coordinates, and $\Ba$ ($\Bz$) is the associated field component, respectively.  
$\Bo$ is the magnetic field strength on the FR axis.
$H = \pm 1$ is the handedness parameter or equivalently the sign of the FR magnetic helicity.
$J_1$ and $J_0$ are the ordinary Bessel functions of order $1$ and $0$.
Finally, $\alpha$ is a parameter related to the amount of twist per unit length along the axis ($= \Ba/(\rho \Bz)$).  In particular, in the vicinity of the FR axis the twist is $\alpha/2$. 

\begin{figure}    %%%%%%%%%%%%%%%%%% FIGURE 1 
  \centering
  \includegraphics[width=\textwidth,clip=]{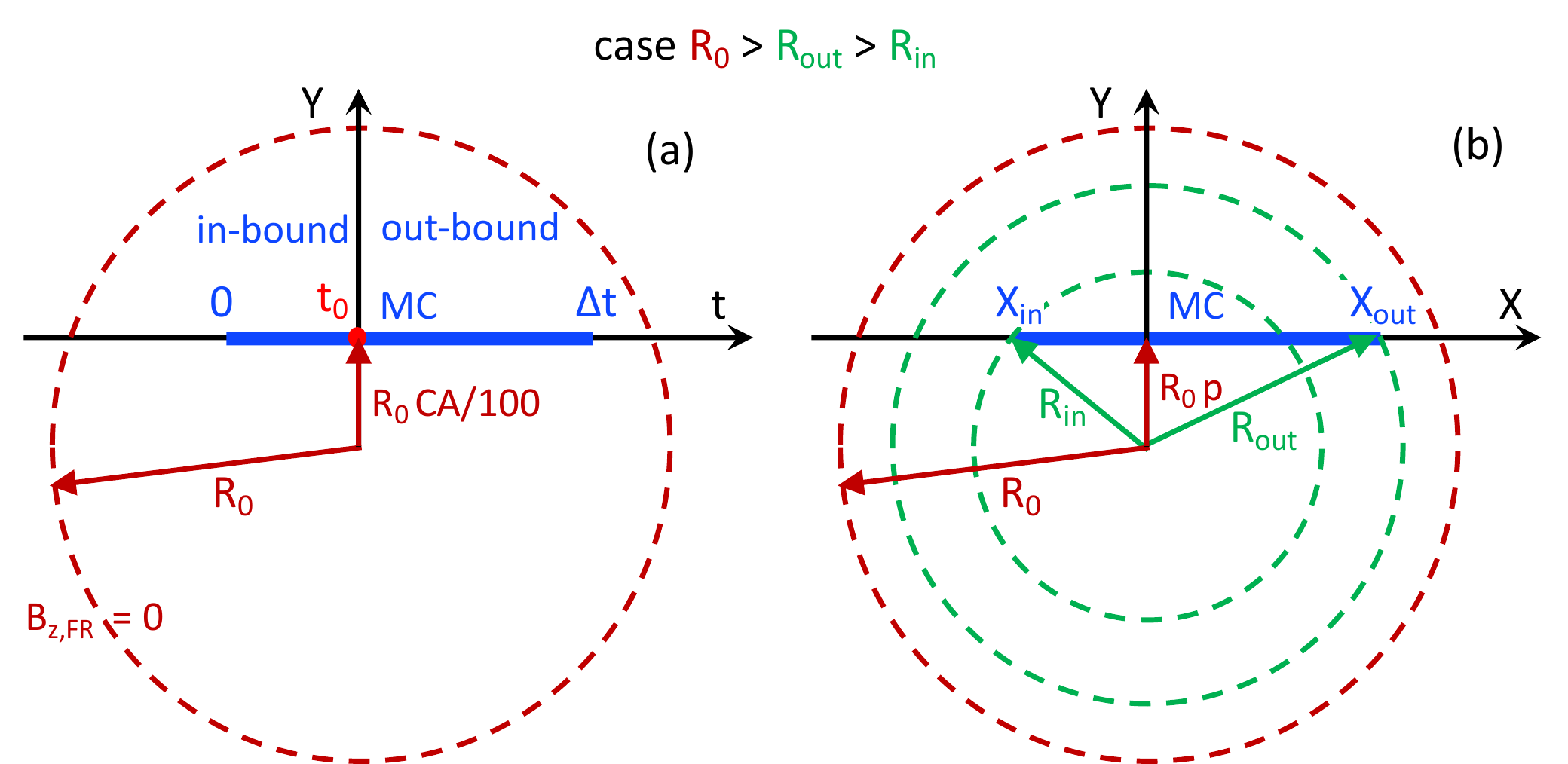} 
  \caption{ Schema defining the parameters used in the cross section of the Lundquist's model fitted to \insitu\ observations. 
    The schema are drawn in the FR cross section with $X$/$Y$ along/across, respectively, the spacecraft trajectory projected in the plane orthogonal to the FR axis (along $Z$). $X,Y,Z$ define the coordinates in the FR frame. 
  The MC extension along the spacecraft trajectory is represented with the blue segment.  The circle with a red dashed line is where the axial field component vanishes.  This defines the radius $\Ro$ (\eq{Ro}).  
    (a) Parameters defined by Lepping's method. The time origin, $t=0$, is set at the first encountered boundary, while the MC duration is $\dt$.  The closest approach of the trajectory from the FR axis is at time $\to$ and at a distance defined by the percentage $CA$ of $\Ro$.
    The region $0<t<\to$ is called the inbound and the region $\to<t<\dt$ is the outbound.
    (b) Parameters defined in this work: the coordinate $X$, the impact parameter $p$,  and radius $R$ at the in- and outbound boundaries.  The case $\Ro>\Rout>\Rin$ is shown while any other ordering is possible with the fit results of MCs observed at 1 au.
          }
  \label{fig_1_notations}
  \end{figure}

    %{\S}{\bf --- Lepping procedure - uniform V} \\
In order to fit $\BL (\rho )$ to the observed components $\Bobs (t)$ within a MC, the time $t$ needs to be converted to the spatial coordinate along the spacecraft trajectory as follows.  The magnetic field is supposed to not evolve during the spacecraft crossing \citep[in particular the FR expansion is not included,][]{Lepping90}.  Then, the time-to-space conversion is modeled by using the mean proton velocity $\Vmean$ across the observed MC.  This defines the spatial coordinate along the spacecraft trajectory with $\Vmean \, t$ and where, for conveniency, the inbound boundary is set at time $t=0$ and the outbound boundary is at $t=\dt$ (\fig{1_notations}a).

    %{\S}{\bf --- Is uniform V justified} \\
Plasma measurements within MCs show that the proton velocity speed is typically decreasing while a spacecraft crosses a MC \citep{Lepping03b, Jian08c, Masias-Meza16}. This indicates that MCs are generally in expansion.  From the observed velocity, the self-similar expansion factor could be derived \citep{Demoulin08,Gulisano10}.  MCs are found to expand at a comparable rate, independently of their size or field strength.   The expansion has an increasing effect on the observed magnetic field with a larger MC duration.  Still, the effects of expansion are moderate even for large MCs \citep[Section 4.3,][]{Demoulin08}.
Then, below we stick with Lepping's method, ignoring expansion.

    %{\S}{\bf --- Lepping procedure - $\Ro$ parameter} \\
   Next, the FR axis orientation is defined by its latitude $\tA$ and longitude $\pA$ (defined in the frame of the observations).  The model also needs to be rescaled in field strength with $\Bo$ and in spatial size with another parameter called $\Ro$.  This radius is defined at the first zero of $J_0$, 
  \begin{equation}  \label{eq_Ro}
  J_0(\alpha \,\Ro ) = 0 \,,
  \end{equation} 
then $\alpha = c/\Ro$, with $c$ the first zero of $J_0$ ($\approx 2.4$).
  It is worth to note that this scaling parameter in size, $\Ro$, could have been defined at any other location of the model, \eg\ where axial and azimuthal components have an equal magnitude.  Different rescaling choices would not affect the fitting results, provide that the spatial rescaling is properly taken into account. 

    %{\S}{\bf --- Lepping procedure - other parameters} \\
The model also needs to have the freedom to be shifted, both along and across the spacecraft trajectory.  This introduces two extra parameters defined when the spacecraft trajectory is at the closest approach from the FR axis.  These parameters are the time $\to$ (or equivalently, the position $\to \, \Vmean$ along the trajectory) and the minimum distance of the spacecraft trajectory to the FR axis. This distance is introduced with the closest approach parameter, CA, as a \% of $\Ro$ (\fig{1_notations}a).

  %{\S}{\bf --- Analysis of the free fitted parameters } \\
Summarizing, six parameters ($\Ro,\Bo,\tA,\pA,\to$ and CA) and a selection parameter ($H=\pm 1$) are involved in the fit.   They are the minimum number of parameters needed for any FR model fitted to the data if no constraint is imposed, such as imposing the same size for the in- and outbound (\eg\ with $\to=\dt/2$) or forcing $\BzFR=0$ at the MC boundaries.
Said differently, the above fit has no intrinsic parameter attached to the Lundquist's model, since $\Bo$ and $\Ro$ are just rescaling parameters in field strength and spatial size, respectively,  while other parameters are geometrical parameters linked with the encounter and the sign of the magnetic helicity.  Finally, the parameter $\alpha$ of \eq{Lundquist} is simply related to $\Ro$ ($\alpha = c/\Ro$) so $\alpha$ is only defined by rescaling the model in size.  Then, there is no extra parameter, \eg\ describing the twist profile, when using the Lundquist's model.  
 
  %{\S}{\bf --- Procedure to fit} \\
The MC magnetic field norm, $B(t)$, is often asymmetric with respect to $t=\to$ as a consequence of both expansion and intrinsic spatial asymmetry (\eg\ compression by the sheath pressure).  To facilitate the fit convergence and to limit the interference of this asymmetry on the deduced FR axis direction, Lepping's method is performed in two steps \citep{Lepping03a}.
  In the first step, the magnetic field vectors from the observations and the model are normalized to unity ($\vec{B}$ changed to $\vec{B}/|\vec{B}|$, then $\Bo = 1$ in this step) and the associated chi-squared, $\chi^2$, is minimized.   $H$ is determined by selecting the lowest $\chi^2$ value given by the two possible signs. The quality of the fit is characterized by the square root of the reduced chi-square defined as $\chiR=\sqrt{\chi^2/(3N_{\rm d}-n)}$, where $n=5$ is the number of parameters of the fit and $N_{\rm d}$ is the number of data points \citep{Lepping10}.
  In the second step, a fit to the original data is performed by minimizing $\chi^2$, this time on the full vectors to determine $\Bo$ (keeping $\Ro,\tA,\pA,\to$ and CA fixed).  Since the fields are not normalized, $\chi^2$ is different from the previous step and its reduced form, $\chiB$, is a measure of the magnetic field difference between the observations and the model.   

  %{\S}{\bf --- Fit quality} \\
    $\chiR$ is used to characterize the quality of the fit, and it is one of the criteria used to associate a fitting quality to MCs \citep{Lepping90}. In particular, $\chiR$ is supposed to quantify the goodness of the FR orientation. However, using the normalized fields does not guarantie that the spatial asymmetry was fully removed (there is no correction along the spacecraft trajectory while magnetic compression implies both an enhanced field and a shorter spatial extention).  Then, we emphasize that $\chiB$ is a better estimation of the fit quality than $\chiR$.   

  %{\S}{\bf --- Quality groups} \\
The MCs are set in three quality groups: good ($Q_1$), fair ($Q_2$), and poor ($Q_3$) as defined in the Appendix A of \citet{Lepping06}. Quality $Q_3$ is first defined with all the MCs with too high values of $\chiR$ or one of the check tests not satisfied.  Next, quality $Q_1$ contains all the MCs with model fit close enough to the data ($\chiR$ below a selected treshold). Finally, quality $Q_2$ contains all the remaining MCs. 

  %{\S}{\bf --- Fit results} \\
Rather than $\to$, the asymmetry factor is provided as a percentage with
   \begin{equation} \label{eq_asf}
   \asf =100 \,|1-2\to/\dt| \, .
   \end{equation}   
 $\asf =0$ when the closest approach of the FR axis is at the center of the observed MC time interval and $\asf =100$ when the closest approach would be at one MC boundary ($t=0$ or $\dt$, \fig{1_notations}).
Inverting \eq{asf} provides $\to$ as
   \begin{equation} \label{eq_to}
   \to = \dt /2 \, (1 \pm \asf /100) \, ,
   \end{equation}   
then the inclusion of an absolute value in the definition of $\asf$ does not permit to distinguish between a closest approach time located before or after the center of the MC time interval.  

The fitted parameters for MCs observed by the \Wind\ spacecraft are available at \url{http://wind.nasa.gov/mfi/mag_cloud_S1.html} \citep{Lepping10}.  

\subsection{New MC Parameters} %%%%%%%%%%%%%%%%%%%%%%%%%%%%
  \label{sect_Fit_New_Par}

  %{\S}{\bf --- Good fitting procedure} \\
The fitting procedure of Lepping described above is fully justified as it does not introduce an apriori knowledge on the FR present within the observed MC, apart that the field is one mode of the linear FFF in a cylindrical configuration.   This description of the magnetic model could be changed to another one, \eg\ Gold and Hoyle's or Vandas's model (see \sect{Intro}).   The same parameters as with the Lepping's method would be present, with extra parameters specific to the magnetic model (\eg\ to describe the cross section shape).  The only difference could be that the radius normalization would need to be changed as the location $\BzFR=0$ may not exist such as in the Gold and Hoyle's model. In this case,  the spatial size is normalized to another radius, \eg\ where azimuthal and axial components are equal.  Another more general normalisation is to set it where the axial field component has decreased by a given fraction of $\Bo$.
  We conclude that the fitting procedure of Lepping is quite general, adaptable to other FR models and built on the strong foundations of the $\chi^{2}$ minimisation.

  %{\S}{\bf --- Explain the problem} \\
While the defined fitting parameters are well justified, and in particular not redundant, their physical meanings are not.  $\Ro$ is defined where $\BzFR=0$ in the fitted model (\eq{Ro}), and this location can be within or outside of the observed MC, with different possible locations in the in- and outbound sides.   There is no justification in Lepping's papers, and subsequent ones, that $\BzFR=0$ should occur or not at both MC boundaries.  This is approximately true for some MCs but false for most of them (\sect{Fit_Radius}). 
   In \sect{Fit_Solar}, we will develop an argumentation to justify $\BzFR=0$ at the FR boundary from the present knowledge of how coronal mass ejections are launched at the Sun.  Still, we also will argue that this ideal case is not expected to be generally present within \insitu\ observations of MCs both because of solar and interplanetary physical processes.   We conclude that the condition $\BzFR=0$ should not be imposed when analyzing the fitted results of Lepping because of the physics involved, and also because the Lepping's fitting procedure does not impose it.   

  %{\S}{\bf --- Transformation of the output parameters} \\
The fitting procedure return a FR model which is defined in the full space with \eq{Lundquist}.  The model is only fitting the data within the observed time interval so it should be considered only in this time interval ($t$ in $[0,\dt ]$, \fig{1_notations}a).  With the conversion of time to the space coordinate $X$ and with $X$ origin set at $t=\to$ (\ie\ at $X$ where the minimum approach distance occurs), the MC boundaries are located at $\Xin$ and $\Xout$ for the in- and outbound, respectively (\fig{1_notations}b).  They are related with the fitted parameters as
  \begin{equation} \label{eq_Xin_Xout}
  \Xin  = -\to \Vmean \cos \lA    \qquad {\rm and} \qquad 
  \Xout = (\dt-\to) \Vmean \cos \lA  \, ,
  \end{equation}
where the location angle $\lA$ is defined as $\arcsin \, (\cos \pA ~\cos \tA )$
\citep[$\lA$ is related to the cone angle $\bA$ as $\lA = 90\degree -\bA$, see][]{Janvier13}. $\cos \lA $ accounts for the projection factor between the spacecraft trajectory and its projection on the FR cross-section (orthogonal to the FR axis). 

  %{\S}{\bf --- Define model radius at MC boundaries} \\
$\Xin$ and $\Xout$ are associated to the model radius $\Rin$ and $\Rout$, defined at the corresponding MC boundaries, as (\fig{1_notations}b)
  \begin{equation} \label{eq_Rin_Rout}
  \Rin  = \sqrt{\Xin^2  +(p\, \Ro)^2}   \qquad {\rm and} \qquad 
  \Rout = \sqrt{\Xout^2 +(p\, \Ro)^2}   \, , 
  \end{equation}
where we introduce the impact parameter, $p$, defined as $p=CA\,/100$.  
Since $\to$ is defined with a sign ambiguity, the results in Lepping's table only allow us to define the inferior and superior values of $R$ at the MC boundaries  \begin{equation} \label{eq_Rinf_Rsup}
  \Rinf = \min (\Xin,\Xout)   \qquad {\rm and} \qquad 
  \Rsup = \max (\Xin,\Xout)   \, ,
  \end{equation}
then, the identification of the in- and outbound is lost.
We avoid labelling these radii with "minimum" and "maximum" since the observed FR could even have a larger radius earlier on during its propagation so that the maximum radius is not necessarily observed due to a possible erosion process \citep[\eg ][]{Dasso06,Lavraud14}.   Finally, it is worth normalizing these radii to $\Ro$ (\eq{Ro}) to better estimate their differences with $\Ro$, 
  \begin{equation} \label{eq_rinf_rsup}
  \rinf = \Rinf / \Ro   \qquad {\rm and} \qquad 
  \rsup = \Rsup / \Ro  \, .
  \end{equation}

  \begin{figure}    %%%%%%%%%%%%%%%%%% FIGURE 2 
  \centering
  \includegraphics[width=\textwidth,clip=]{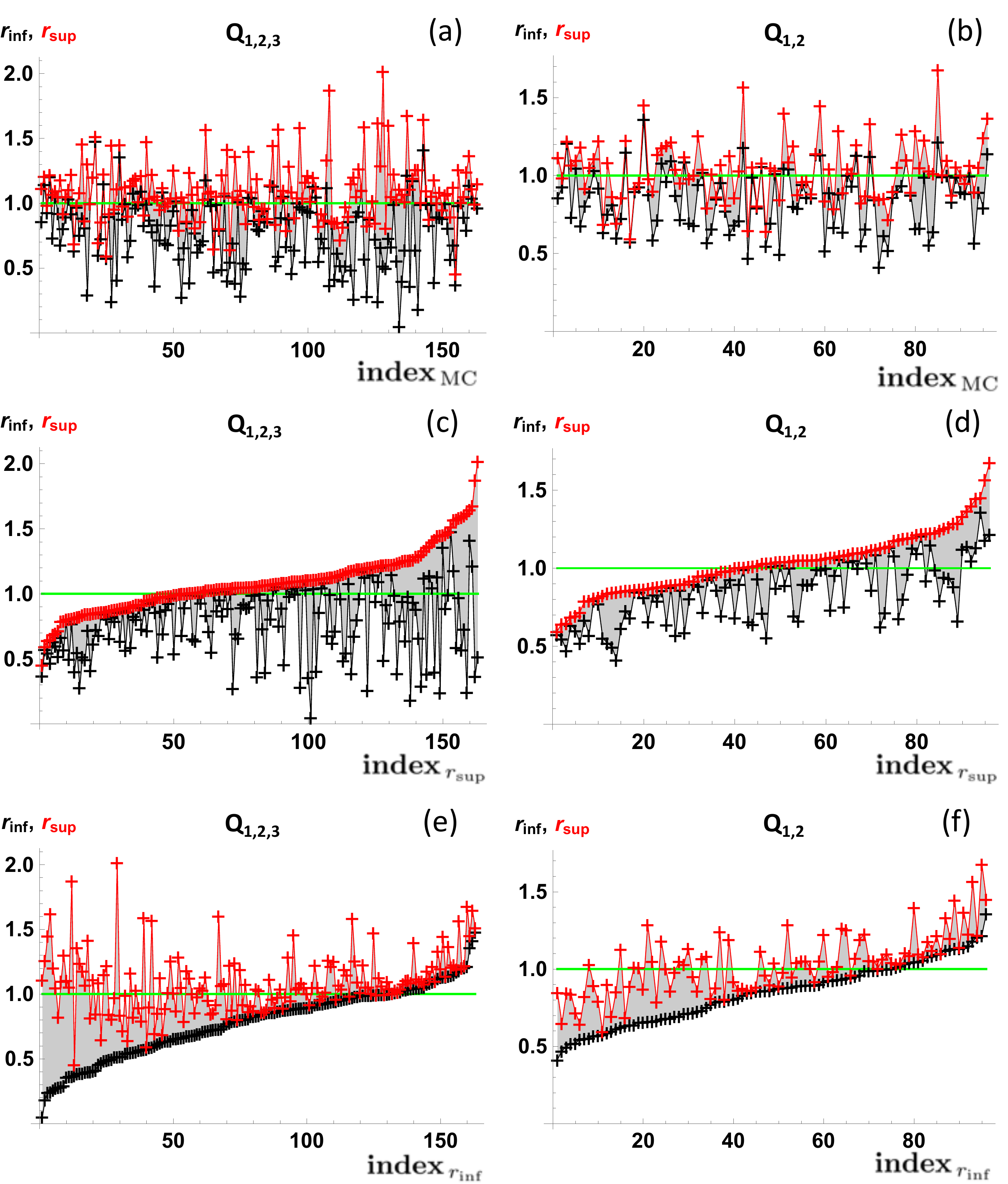} 
  \caption{ Radius, $\rinf$ and $\rsup$, of the fitted FR at the MC boundaries, defined with $\Rinf$ and $\Rsup$, \eq{Rinf_Rsup}, and normalized by $\Ro$ defined by $\BzFR=0$. 
The perfect match of $\Rinf$ and $\Rsup$ with $\Ro$ is marked by an horizontal green line.
    (a,b) The results are ordered with the ranking index of MCs as ordered in Lepping's table (so with observation time). 
    Next, the results are ordered with the index derived from ranking (c,d) $\rsup$ and (e,f) $\rinf$ with growing values. 
    The left column (a,c,e) contains all MCs observed by \Wind\ at 1 au from February 1995 to December 2012 (163 MCs), while the right column (b,d,f) contain only MCs of quality 1 and 2 (96 MCs).
          }
  \label{fig_2_r}
  \end{figure}

  \begin{figure}    %%%%%%%%%%%%%%%%%% FIGURE 3 
  \centering
  \includegraphics[width=\textwidth,clip=]{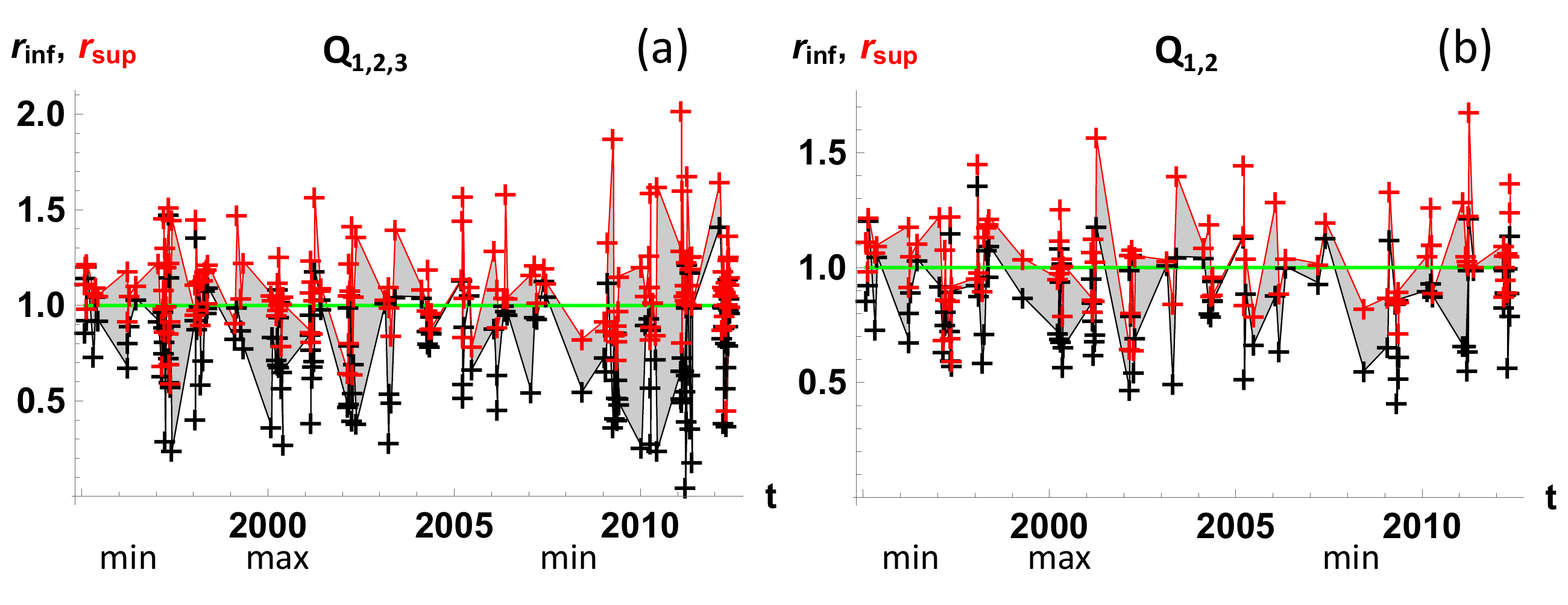} 
  \caption{ Radius, $\rinf$ and $\rsup$, of the fitted FR at the MC boundaries, as in \fig{2_r},
in function of the observed time.  The times of the minimum and maximum of the solar cycle are added at the bottom.  (a) All MCs observed by \Wind\ at 1 au from February 1995 to December 2012, (b) MCs of quality 1 and 2. }
  \label{fig_3_cycle}
  \end{figure}

\subsection{Radius of Observed MCs} %%%%%%%%%%%%%%%%%%%%%%%%%%%%
  \label{sect_Fit_Radius}
The results of \eqs{Rinf_Rsup}{rinf_rsup} applied to Lepping's list are shown in  \figS{2_r}a,b with the MCs ranked by time.  The variable, index$_{\rm MC}$, in abscissa is simply the integer giving the position in the ranked list. This index is almost the event identifier within the Lepping's table except that the identifier has some real values (due to the inclusion of more MCs during the revision of the table).  This index is growing in time from February 1995 to December 2012. 

  %{\S}{\bf --- $\rinf$ and $\rsup$ results with index$_{\rm MC}$} \\
   First, there are significant deviations of both $\rinf$ and $\rsup$ away from 1, so from the ideal FR with $\BzFR=0$ set at the FR boundary (\fig{2_r}a).  These deviations from unity are only slightly reduced when the MC set is restricted to qualities 1 and 2 (\fig{2_r}b).  Moreover, there is no coherence of the deviation between neighbouring MCs, so with time, and in particular no link with the solar cycle (\fig{3_cycle}).  Indeed, the frequency of MCs is function of the year within the solar cycle, then changing the abscissa from index$_{\rm MC}$ to time only regroup the MCs closer around the maximum of the solar cycle, and no coherent variation of $\rinf$ and $\rsup$ is present with the solar cycle.  
   
\begin{figure}    %%%%%%%%%%%%%%%%%% FIGURE 4 
  \centering
  \includegraphics[width=\textwidth,clip=]{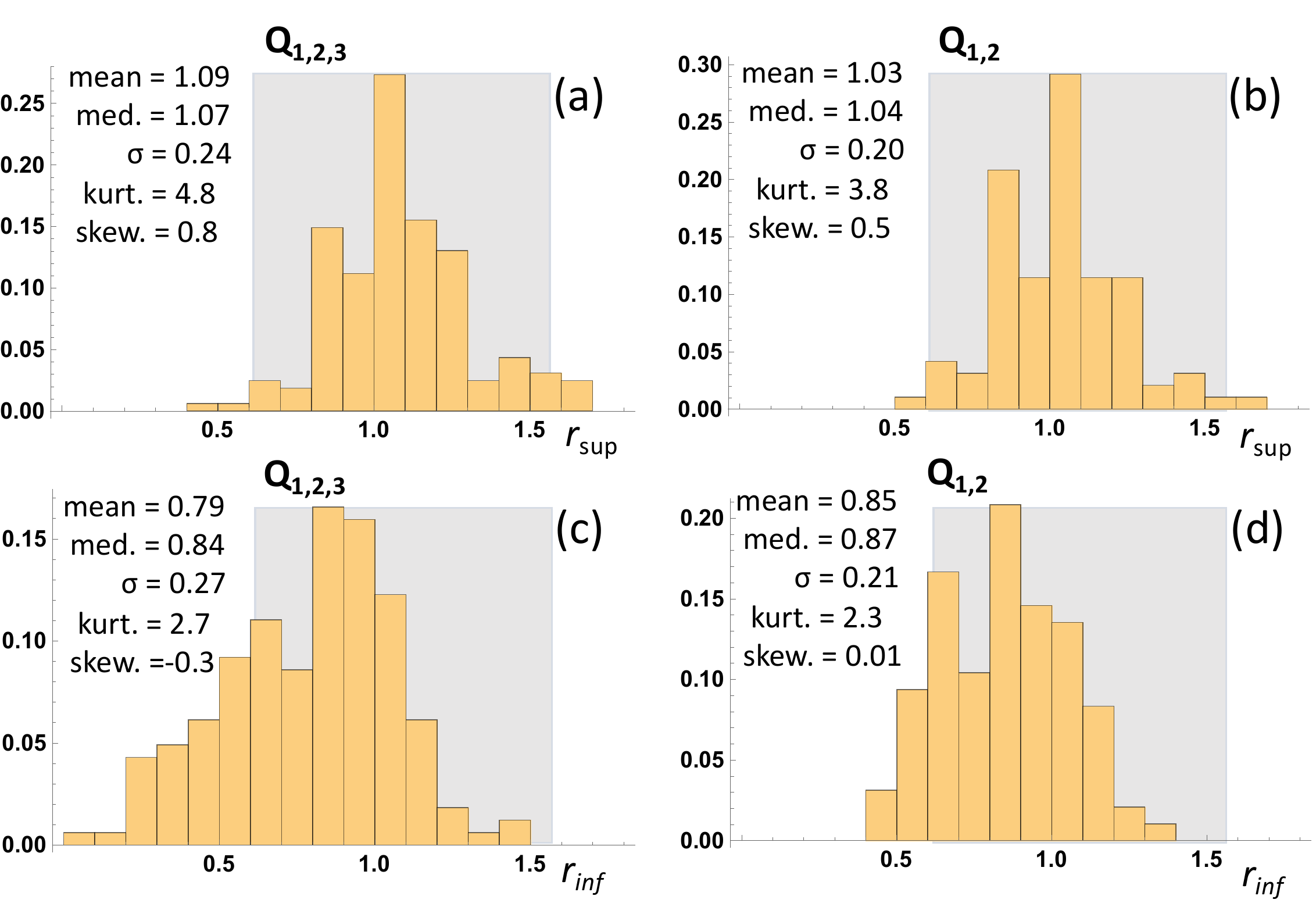} 
  \caption{ Probability distributions of the normalized radius at MC boundaries ($\rinf$ and $\rsup$ defined by \eq{rinf_rsup}) for all MCs (left column) and for MCs of quality 1 and 2 (right column). The horizontal extension of the grey region behind histograms is fixed for all panels to the mean $\pm 2 \sigma$ of panel (a) as a guide for comparison between panels and later on it is used in \figs{histo_fa_fz}{H}.
          }
  \label{fig_4_histo_r}
  \end{figure}

  %{\S}{\bf --- $\rsup$ results with index$_{\rm \rsup}$} \\
  The values of $\rsup$ (in red) are dispersed around unity with only a slightly larger range when quality 3 is included (\figS{2_r}a,b).  We next change the abscissa to a ranking by growing order of $\rsup$ where the index of $\rsup$ list is index$_{\rm \rsup}$.  This orders the results and then allows us to better see the variety and also the organisation of the observed cases (\figS{2_r}c,d).  About 60\% of the MCs have $\rsup>1$, so an axial field reversal with $\BzFR<0$, near at least one of the MC boundaries.  Next, $\rsup$ has almost a linear trend. This is not a trivial result since only a monotonous behaviour is imposed by using index$_{\rm \rsup}$ in abscissa.  More precisely, a steeper slope is present at both ends of the $\rsup$ index range.  
The number of MCs present in a given $\rsup$ interval, \eg\ $\Delta \rsup = 0.1$, is inversely proportional to the slope of $\rsup$(index$_{\rm \rsup}$) shown in \figS{2_r}c,d.  Then, a steeper slope indicates few cases in the wings of the $\rsup$ probability distribution as confirmed by \figS{4_histo_r}a,b (with bin width $\Delta \rsup = 0.1$).   Moreover, it is remarkable that the probability distribution of $\rsup$ values is almost the same when MCs of quality 3 (67 MCs) are added to those of quality 1 and 2 (96 MCs), showing that these supposed worse fitted cases have in fact a similar $\rsup$ distribution.

  %{\S}{\bf --- $\rinf$ results with index$_{\rm rinf}$} \\s
The variety of cases for $\rinf$ and its organisation in the MC data set are better shown with an abscissa defined by the ranking by growing order of $\rinf$, so by introducing the corresponding index$_{\rm rinf}$ (\figS{2_r}e,f).
Even more than $\rsup$,  $\rinf$ is almost a linear function of index$_{\rm \rinf}$.
There is again only a steepening of the slope for few cases at both ends of index$_{\rm rinf}$.   Then, \figS{2_r}e,f implies that the probability distribution of $\rinf$ is relatively flat with fewer cases for the extreme values of $\rinf$.  This is confirmed with an histogram representation (\figS{4_histo_r}c,d). 

  %{\S}{\bf --- $\rsup-\rinf$ with index$_{\rm rinf}$} \\
  The grey region between $\rinf$ and $\rsup$ in \fig{2_r} shows the asymmetry in size of the in- and outbound regions.  While it is present in all the plots of \fig{2_r}, this asymmetry is better seen with $\rinf$ index in abscissa.  For quality 1 and 2 (\fig{2_r}f) the difference is spread between 0 and 0.5 nearly uniformly with $\rinf$ (or $\rsup$) index, while adding quality 3 (\fig{2_r}e) includes many cases which are more asymmetric especially for lower $\rinf$ values.  This is the consequence of one of Lepping's criteria setting some MCs to quality 3 as follows.  In parallel of the fit results, the FR radius is estimated from the fitted parameters by forcing $\to =0$ \citep[Equation 8 in][]{Lepping06}.  When this derived radius is significantly different than $\Ro$, the corresponding MC is set in quality 3 whatever the fit quality $\chiR$ is.   This is not a consistency check as claimed, but rather a filter which removes from qualities 1 and 2 MCs with too large $\to /\dt$ values, so implicitly with a too large $\rsup /\rinf$ ratio (as deduced from \eqss{Xin_Xout}{rinf_rsup}).  This is why \fig{2_r}e has many MCs with a large $\rsup /\rinf$ ratio in contrast of \fig{2_r}f.     

  %{\S}{\bf --- annulus cases: outsiders?} \\
The cases with both $\rinf >1$ and $\rsup >1$ were called annulus cases in Lepping's papers because their FR core ($\BzFR >0$ by definition) is surrounded by an annulus of $\BzFR <0$ \citep{Lepping03a,Lepping06}.
These MCs also were described as special cases while \figS{2_r}e,f show that they are in fact only part of a quasi-continuous distribution (larger values of $\rinf$).  Moreover the inclusion of quality 3 (\fig{2_r}e) does not change significantly the right side of the plot compared with quality 1 and 2 (\fig{2_r}f), so the relative number of annulus cases, while it does on the small radius side (due to the selection describe above). About 20\% of cases have $\rinf>1$, so $\BzFR<0$, at both MC boundaries.  We conclude that these annulus cases are not outsiders or MCs perturbed / not well observed, but they are only a part of the distribution of FR cases launched from the Sun. 

  %{\S}{\bf --- $\rsup \,{\rm and}\, \rinf$ distributions} \\
Next, the distributions of $\rsup$ and $\rinf$ are shown in \fig{4_histo_r}.  $\rsup$ distributions are nearly centered on $r=1$ (both mean and median slightly above $1$) and approximately symmetric (small skewness).  They also have a narrow core with small wings (large kurtosis).  
  In contrast the distribution of $\rinf$, \figS{4_histo_r}c,d, are centered at values around $r=0.8$.  They are flatter (lower kurtosis than \figS{4_histo_r}a,b) while still approximately symmetric (low skewness).  Adding the cases of quality 3 increases the extension to lower $r$ values, then increases slightly the standard deviation $\sigma$ of $\rinf$ distribution (due to the selection of MCs as explained two paragraphs above).

  \begin{figure}    %%%%%%%%%%%%%%%%%% FIGURE 5 
  \centering
  \includegraphics[width=\textwidth,clip=]{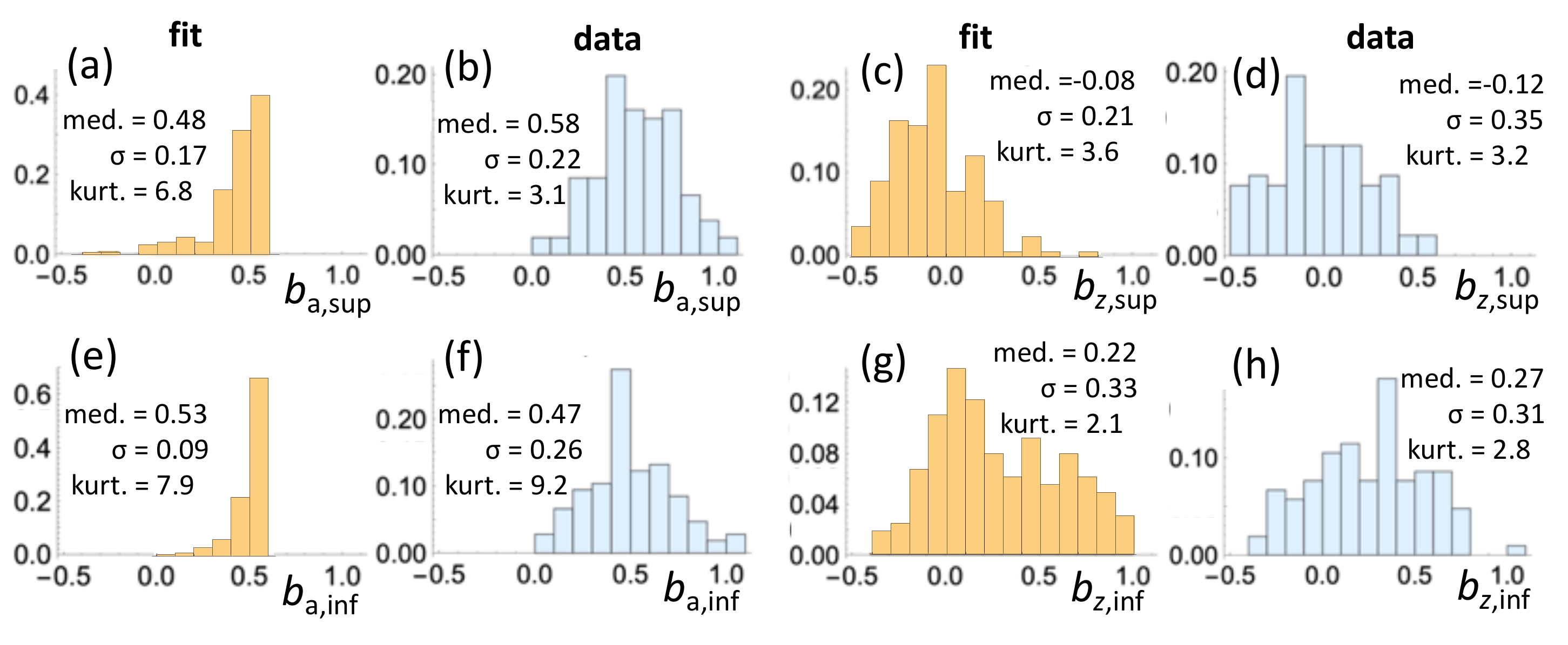} 
  \caption{Probability distributions of the azimuthal $\ba =\Ba/\Bo$ (left half side) and axial $\bz =\Bz/\Bo$  (right half side) magnetic field components at the MC boundaries associated with $\rsup$ (top row) and $\rinf$ (bottom row).  
  All components are normalized by the central field magnitude $\Bo$.
  In panels with orange histograms (a,c,e,g) the components are computed from Lundquist's fit while in panels with blue histograms (b,d,f,h) the components are from the \insitu\ data rotated in the FR frame. All the MCs of the data set are used (quality 1-3).
          }
  \label{fig_histo_ba_bz}
  \end{figure}
 
\subsection{Field Components at MC Boundaries} %%%%%%%%%%%%%%%%%%%%%%%%%%%%
  \label{sect_Fit_B}

  %{\S}{\bf --- fit: $\BzFR=0$ in general not satisfied at MC boundaries} \\
The analysis above showed that the FR radii obtained from observations of the MC boundaries are frequently different than $\Ro$ (\eq{Ro}). This is equivalent to say that $\BzFR \neq 0$ in general. This can be seen directly on the right half of \fig{histo_ba_bz} with the histograms of $\bz =\Bz /\Bo$ at the MC boundaries.  The results with the Lundquist fit applied to all MCs is shown in \figS{histo_ba_bz}c,g.  $\bzsup$ is nearly centered on 0 (mean $=-0.06$, median $=-0.08$) but with a broad distribution ($\sigma=0.21$), while $\bzinf$ is centered on $\approx 0.25$ (mean $=0.28$, median $=0.22$) and with a broader distribution ($\sigma=0.33$).  These results only weakly change if MCs are restricted to qualities 1 and 2 (the main differences are $\bzinf$ centered on $0.19$ with a smaller $\sigma$ of $0.26$). 

  %{\S}{\bf --- observations: $\BzFR=0$ in general not satisfied at MC boundaries} \\
The value of $\Bz$ at the MC boundaries also can be analysed directly from the observations after rotating them in the FR frame.  The associated histograms are shown \figS{histo_ba_bz}d,h with the same normalization by $\Bo$.  These histograms have differences with respect to the ones derived from the Lundquist fit.  Indeed, the data values are more sensitive to the local fluctuations of the magnetic field and the precise definition of the MC boundaries.   Still, the data histograms of \figS{histo_ba_bz}d,h are comparable if drawn slightly within the MC or restricting MCs to qualities 1 and 2 (not shown).   
  Moreover, the main point is that the histograms derived from the Lundquist fit and the data are globally comparable, leading to the same conclusion that $\BzFR=0$ is typically not well satisfied at the MC boundaries, and especially on one side (corresponding to $\rinf$).  More precisely, $\bzsup<0$ is a bit more present in the data than in the fit results, while $\bzinf<0$ is present for a comparable number of cases, so the analysis of data show a similar proportion of annulus cases.

  %{\S}{\bf --- $\Ba$ values at MC boundaries } \\
 The same analysis is performed with the azimuthal field component at the MC boundaries. Since the function $\ba (r)=J_{1}(r)$ is maximum for $r \approx 0.77$, which is about the mean value of $\rinf$ (\figS{4_histo_r}c), the histogram of $\bainf$ is clustered below the maximum value of $J_{1}$ ($\approx 0.58$, \fig{histo_ba_bz}e).  A similar clustering of $\basup$ is present (\fig{histo_ba_bz}a), while less important since $\rsup$ is centered near 1 (\figS{4_histo_r}a). 
Then, with the Lundquist fit, the values of $\ba$ at MC boundaries are much less spread than those of the $\bz$ component.
  In contrast, the histograms of $\ba$ derived directly from the data have a significant tail with $\ba$ up to unity (\figS{histo_ba_bz}b,f).  This implies that the Lundquist fit typically provides a too low azimuthal field near the FR boundaries.

  %{\S}{\bf --- Interpretation of $r$ and $\bz$ distributions} \\
  The distributions of $r,\ba,\bz$ (\figs{4_histo_r}{histo_ba_bz}) can be understood in the context of FR erosion during the interplanetary travel \citep{Dasso06,Ruffenach15}.  
   Supposing that FRs are formed in the corona with $\bz \approx 0$ at their boundaries (\sect{Fit_Solar}), the distributions of $\rsup$ and $\bzsup$ centered on 1 and 0, respectively, are compatible with FRs that are not significantly eroded on one side.  With significant erosion they would be more similar to the distributions of $\rinf$ and $\bzinf$. Indeed, this is an expected result since fast MCs are expected to be eroded at the front (reconnection with the overtaken solar wind magnetic field which is forming the sheath magnetic field), while slow MCs are rather expected to be eroded at the rear when they are overtaken by a fast stream.  The amount of erosion is variable depending on the forcing (\eg\ difference of velocities and relative orientation of the magnetic fields).  More generally, we cannot insure if $\rsup$ and $\bzsup$ kept the solar distributions or are also transformed by erosion. 
    In contrast, the larger extension of the distributions of $\rinf$/ $\bzinf$
towards smaller/larger values, respectively, is characteristic of erosion or/and that FRs were never fully formed in the corona but get an arcade like flux on the front.

  %{\S}{\bf --- Conclude} \\
   We conclude that, because of the physical processes involved, it is important to not impose $\BzFR=0$ at the MC boundaries when fitting a FR model to the \insitu\ data.

%________________________________________________________________________
\begin{figure}[t!]       %%%%%%%%%% FIGURE 6 Correlations 
\centering
 \includegraphics[width=0.9\textwidth, clip=]{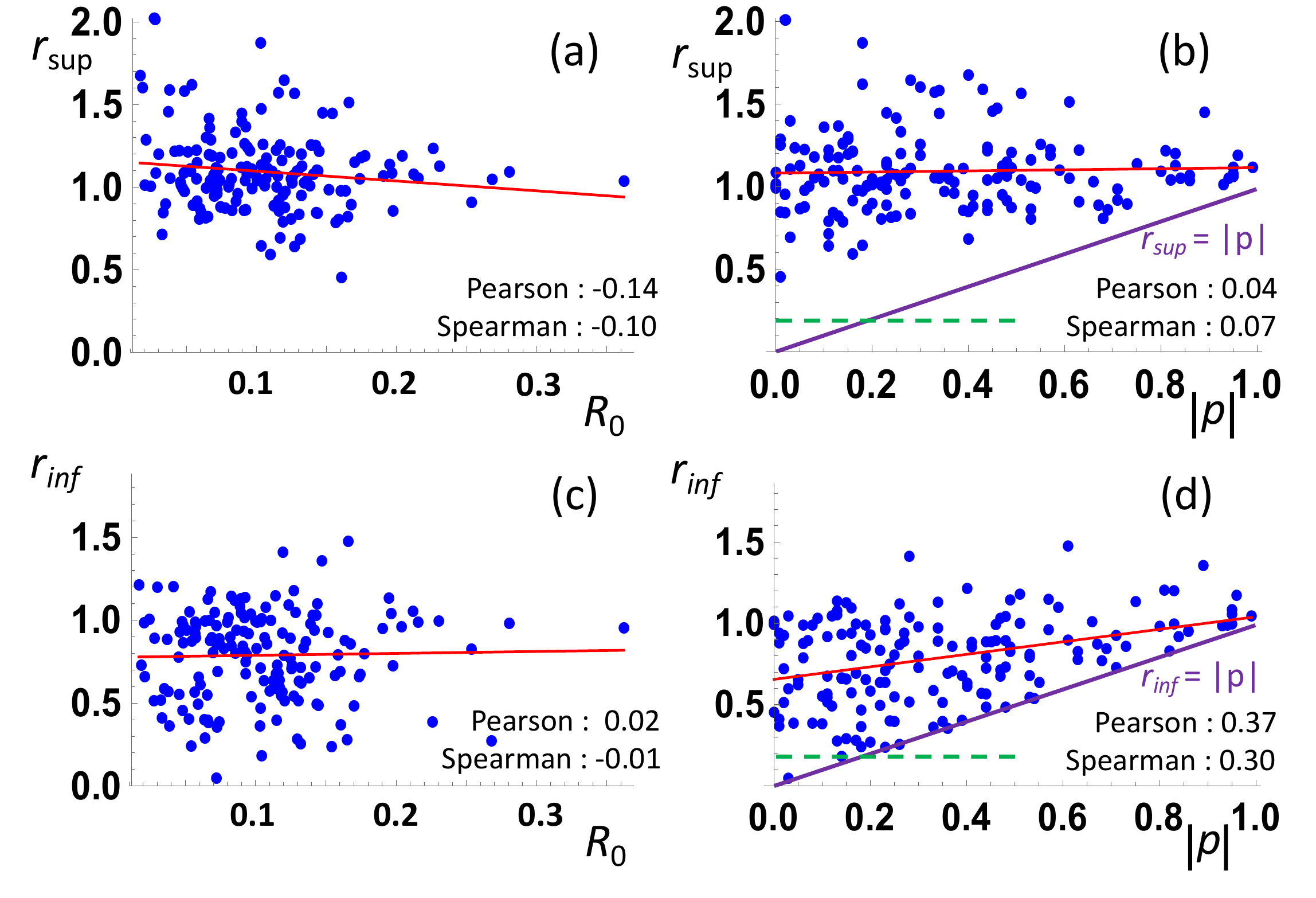}
\caption{Examples of correlation analysis of $\rsup$ (top) and $\rinf$ (bottom) for the full set of MCs. The red lines are least square fits with a linear function to derive the main tendency. The Pearson and Spearman correlation coefficients are reported.
  (a,c) Correlations with $\Ro$.
  (b,d) Correlations with $|p|$.  The purple line in (d), $\rinf=p$, is the limit of detection of a FR on one side (either in- or outbound). The dashed green line indicates a plausible bias due to the non detection of too asymmetric FRs in terms of in- or outbound extensions.
}
 \label{fig_corel}
\end{figure}
\subsection{Correlations with MC Physical Parameters} %%%%%%%%%%%%%%%%%%%%%%%%%%%%
  \label{sect_Fit_Correlations}

  %{\S}{\bf --- Absence of correlations} \\
We test whether $\rinf$ and $\rsup$ are correlated with global MC physical parameters by performing a least square fit of a linear function in order to derive the main tendency, and by computing the Pearson and Spearman (or rank) correlation coefficients.  Examples of results are shown in \figS{corel}a,c for $\Ro$.  
No correlation is found for $\Ro$, $\Vmean$, $\tA$ and $\pA$ as the absolute value of the correlation coefficients is close to or below $0.1$ for all these parameters.  A weak positive correlation is found with $\Bo$, with correlation factors between 0.14 and 0.28 but these correlations are tied to few (4 or 5) outsider cases with extreme $\Bo$ values ($\geq 40$ nT). 
 
  %{\S}{\bf --- Correlation with $|p|$} \\
The only significant correlations found are between $\rinf$ and the absolute value of the impact parameter $|p|$ (\fig{corel}d).  However, the condition $\rinf > |p|$ is needed to detect the smallest FR side.  For example, in \fig{1_notations}b, where $\rinf=r_{\rm in}$ was selected, the limiting case corresponds to $\Xin =0$, then $\Rin = |p| \, \Ro $ so that the spacecraft cannot observe the inbound region.  This condition is well present in \fig{corel}d with only data points above this limit (purple line).   This condition is at the origin of most of the correlation of $\rinf$ with $|p|$ since only a weak increase of $\rinf$ with $|p|$ is present at the top part of $\rinf$ values.  Since $\rsup$ is frequently well above $\rinf$, the condition $\rsup > |p|$ affects only weakly $\rsup$ (at most only for the largest $|p|$ values) and $\rsup$ is uncorrelated with $|p|$ (\fig{corel}b).

  %{\S}{\bf --- Other possible bias of $\rinf$} \\
Another observational selection is likely to be present for $\rinf$ with no data points below the dashed green line (\fig{corel}d). Then, only MCs with large enough in- and outbound, $\rinf>0.2$, then not too eroded, are defined as MCs.  An equivalent observational selection has no effect on $\rsup$ (\fig{corel}b). 
This apparent selection effect on $\rinf$, for $|p|<0.2$, is still surprising since for larger $|p|$ values some MCs (approximately 15 cases) are observed with $\rinf \approx |p|$ (blue points close to the purple line in \fig{corel}d).  This corresponds to about 10\% of observed MCs defined with no in- or outbound crossed (so only less than half of the FR crossed by the spacecraft).

  %{\S}{\bf --- Conclusion} \\
We conclude that $\rinf$ and $\rsup$, as well as their ratio, have no significant correlation with the global physical parameters and those describing the encounter geometry when estimated for the full set of MCs.   The same conclusion is reached when restricting the MCs to qualities 1 and 2 (not shown). These results indicate that the amount of erosion is independent of the MC properties.
 
\subsection{Solar Origin of Flux Ropes} %%%%%%%%%%%%%%%%%%%%%%%%%%%%%
  \label{sect_Fit_Solar}      

  %{\S}{\bf --- Generalities on eruptive FRs} \\
The solar origin of MCs is typically eruptive flares in active regions or eruptions of quiescent filaments.  These flares have typically two ribbons which trace the chromospheric feet of the field lines (FLs) involved in the magnetic reconnection. A first set of reconnected FLs are observed as flare loops.  A second set is observed as sigmoidal loops which expand and move up rapidly.  They trace the erupting FR which is further enlarged with the new reconnected flux wrapped around it as the FR is ejected upward \citep[\eg ][and references therein]{Janvier17,Welsch18}.   

\begin{figure}[t!]       %%%%%%%%%% FIGURE 7. Flare evolution 
\centering
 \includegraphics[width=0.9\textwidth, clip=]{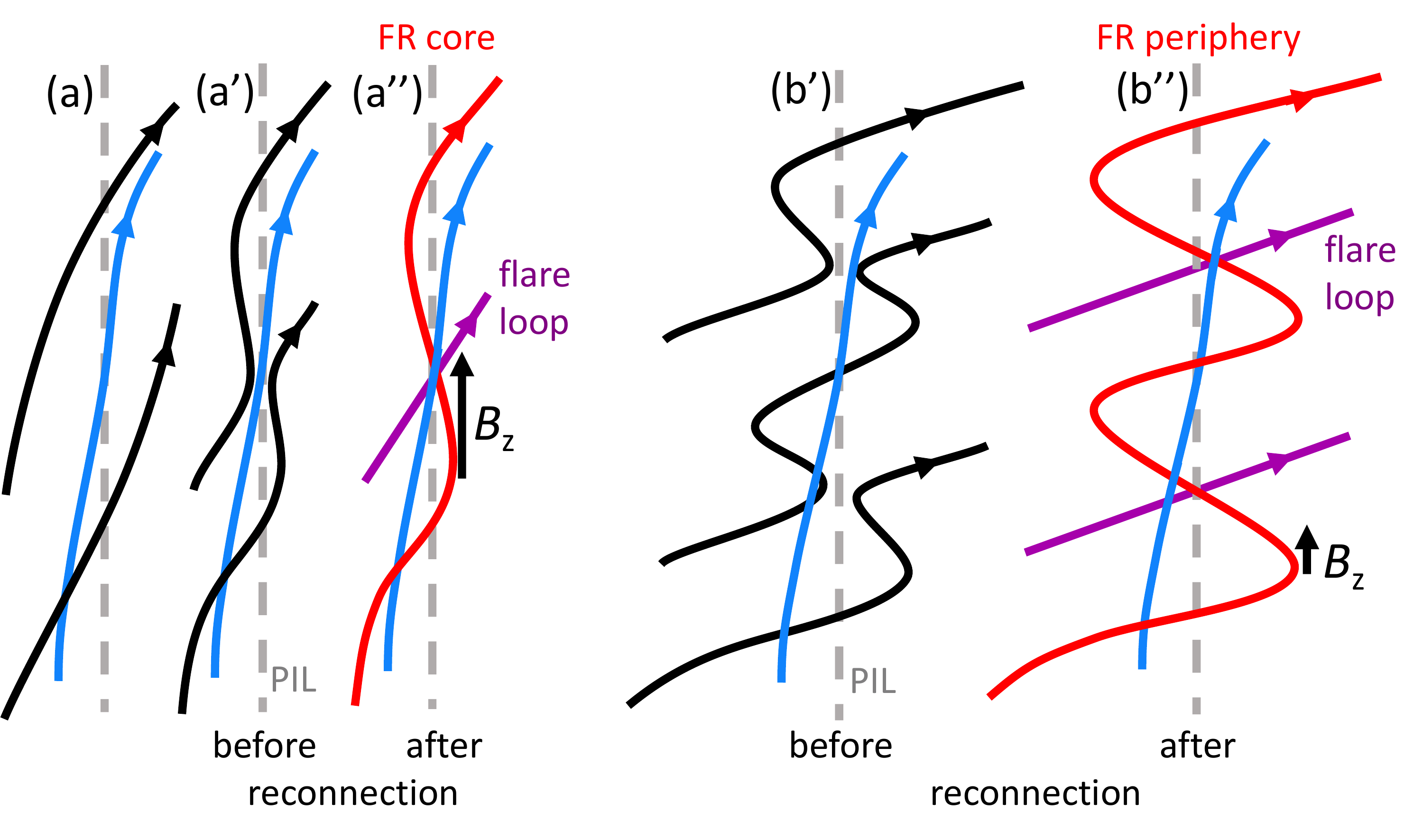}
\caption{ Evolution of magnetic configuration for an eruptive flare or for a quiescent filament eruption as seen from a top view.  The system of sheared arcades (black) is progressively transformed to a FR (red) and an underlying arcade of flare loops (purple) during the eruption. (a) shows two FLs (black) which are progressively brought nearby below the FR core (outlined by a blue FL) in panel (a$^{\prime}$).    Next, two reconnection steps are shown.  
  (a$^{\prime}$,a$^{\prime \prime}$) shows a step at the flare beginning, which enlarges the FR core. 
  (b$^{\prime}$,b$^{\prime \prime}$) shows another step later in the flare (we omit the earlier step similar to panel (a)), which build up the FR periphery (the central black FL reconnect twice, in general at different times). Because the original arcade (in black) is more sheared closer to the photospheric inversion line, a less twisted core than the periphery is built by this reconnection process (corresponding to a stronger $\Bz$ field component in the FR core).
}
 \label{fig_flare}
\end{figure}

  %{\S}{\bf --- Shear evolution of flare loops} \\ 
A direct deduction from the coronal images of the FR structure, and in particular its twist profile, is typically difficult as the bright loops tracing the FR are usually 
only a few, not well observed all along the FR, and they have rapidly a too weak brightness compare to the background as the FR expands and is ejected.   Flare loops below, being less extended and in a more dense region, are typically much better observed all along the flare duration.  The first formed loops have typically a magnetic shear. Then, as higher flare loops are formed below the erupting FR, they become more potential, \ie\ more orthogonal to the underlying photospheric inversion line (PIL) of the vertical component of the magnetic field \citep[\eg ][]{Asai03,Su06,Liu10}.  A devoted study confirms these findings both observationally and in a numerical model of eruptive flares \citep[Figures 2 and 7, respectively, of][]{Aulanier12}.   In parallel to this shear evolution, the successive layers added to the eruptive FR are formed by more twisted field lines, as summarized in \fig{flare}.   

%________________________________________________________________________

   %{\S}{\bf --- $\Bz \approx 0$ at FR boundary} \\
In the idealized case described above, the erupting FR is progressively wrapped by the surrounding arcade which is formed by a more potential field at large heights.   Supposing that this wrapped field could reconnect behind the erupting FR, this implies a negligible axial field at the FR periphery.   This is a justification for selecting $\BzFR = 0$ at the FR boundary of a model fitting \insitu\ MC data.   However, solar configurations are more complex than this simple description. For example, the PIL is rarely straight along the eruptive FR, and part of the upper arcade could belong to another magnetic bipole \citep[\eg ][]{Schrijver11,vanDriel14,vanDriel15}.  This implies that the potential field may be not orthogonal to the local PIL.  Furthermore, the potentiality of the arcade is not guaranteed; it can even have an opposite helicity sign than the erupting FR \citep[\eg ][]{Cho13,Vemareddy17}.  Finally, erupting FR in simulations are frequently observed to rotate \citep[with transfer of twist to writhe helicity, \eg][]{Kliem12}.  This introduces another variability for the $\BzFR$ component of the overlying/reconnecting arcade.

   %{\S}{\bf --- Conclusion} \\
We conclude that from the coronal physics of the eruption, a distribution of cases around the condition $\Bz = 0$ at the FR boundary is to be expected.   Furthermore, reconnection at the front or at the rear of the FR could happen during the travel from the Sun when magnetic fields of different origins, so different orientations, are pushed again each others by the difference of velocity (between the MC front and the sheath or/and between the MC rear and an overtaking fast stream or another event).   Then, when analyzing \insitu\ data the condition $\Bz = 0$ should not be imposed at the MC boundaries.

  \begin{figure}    %%%%%%%%%%%%%%%%%% FIGURE 8  fa fz
  \centering
  \includegraphics[width=\textwidth,clip=]{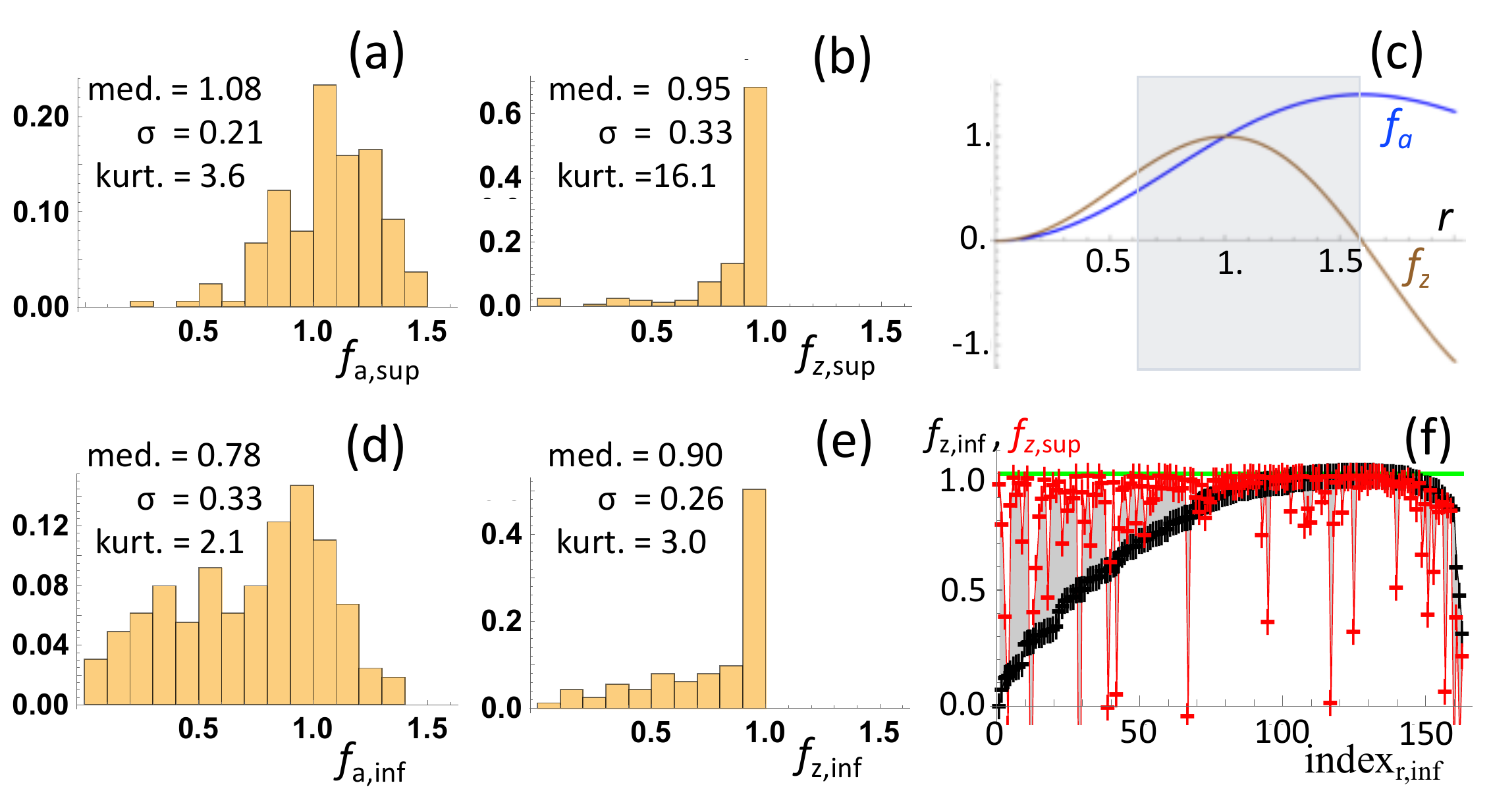} 
  \caption{ Magnetic fluxes computed with all MCs (Q$_{1,2,3}$) and the fitted Lundquist model.
    (a,d) Azimuthal flux $\fa$ and (b,e) axial flux $\fz$ histograms.  These fluxes are computed with a FR model extending up to (a,b) $\rsup$ or to (d,e) $\rinf$ and they are normalized to 1 for $r=1$.   
   (c) Behavior of $\fa (r)$ and $\fz (r)$ in function of the radius $r$ (normalized to $\Ro$).  The horizontal extension of the grey region is defined by the mean $\pm 2 \sigma$ for $\rsup$ and Q$_{1,2,3}$ (defined in \fig{4_histo_r}a).
   (f) Axial flux associated to FRs with radius $\rinf$ and $\rsup$. The results are ordered with growing values of $\rinf$ (5 MCs have $\fzsup<-0.1$, so they are located below the plot limits). 
          }
  \label{fig_histo_fa_fz}
  \end{figure}
  
%%%%%%%%%%%%%%%%%%%%%%%%%%%%%%%%%%%%%%%%%%%%%%%%%%%%%%%%%%%%%%%%%%%%%%%%%%%%%%%%%%
\section{Implications} %%%%%%%%%%%%%%%%%%%%%%%%%%%%%%%%%%%%%%%%%%%%
  \label{sect_Implications}

\subsection{Magnetic Flux Imbalance} %%%%%%%%%%%%%%%%%%%%%%%%%%%%%
  \label{sect_Implications_Flux}      

   %{\S}{\bf --- Magnetic flux functions} \\
The azimuthal and axial fluxes, within a radius $\rho$ of a cylindrical FR, are simply computed by integration of \eq{Lundquist} with the inclusion of the geometrical factors,
  \begin{equation} \label{eq_Fa_Fz}
  \Fa (\rho) = L \Int{0}{\rho} \Ba (\rho') \, \rmd \rho'     \quad {\rm and} \quad 
  \Fz (\rho) = 2\pi \Int{0}{\rho} \rho' \Bz (\rho') \, \rmd \rho' \,, 
  \end{equation}
where $L$ is the length considered along the FR axis. 
After normalization by the corresponding fluxes obtained for the radius $\Ro$ (\eq{Ro}), then at  $r=\rho/\Ro =1$, the normalized fluxes $\fa (r) =\Fa (\rho)/\Fa (\Ro)$ and  $\fz (r) =\Fz (\rho)/\Fz (\Ro)$ write (using $J_0(\alpha)=0$)
  \begin{equation} \label{eq_fa_fz}
  \fa (r) = 1 - J_0(\alpha r)     \qquad {\rm and} \qquad 
  \fz (r) = r\, J_1(\alpha r) \,/J_1(\alpha)\,.  
  \end{equation}
They are plotted in \fig{histo_fa_fz}c.
   
   %{\S}{\bf --- $\fa$ spread} \\
Since $\fa (r)$ is approximaly linear in the interval $\pm 2 \sigma$ around the mean of $\rsup$ for all MCs (grey band in \figs{4_histo_r}{histo_fa_fz}c), the distribution of $\fasup$ in \fig{histo_fa_fz}a is similar to the distribution of $\rsup$ in \fig{4_histo_r}a.  
The largest difference is a distribution of $\fasup$ less extended to large values than with $\rsup$ because $\fa (r)$ is maximum near $\approx 1.6$.
Next, as $\rinf$ is smaller and more spread than $\rsup$ (\fig{4_histo_r}c), the non-linearity of $\fa (r)$ implies an even broader distribution of $\fainf$ towards lower values (\fig{histo_fa_fz}d) than the distribution of $\rinf$ (\fig{4_histo_r}c).  Then, the large dispersion of $r$ values at the MC boundaries strongly affects the determined azimuthal magnetic flux so that it cannot be approximated by its value obtained with $\Ro$, at least for individual MC studies.   This difference could be minimised with a statistical study, especially for $\fasup$ which has a distribution centred close to 1 (\fig{histo_fa_fz}d).  Furthermore, a statistical study which incorporates MCs with a broad range of flux $\Fa$ further minimises this difference of $\fa$ values.  For example, \citet{Qiu07} and \citet{Gopalswamy17} were able to derive an important correlation between the azimuthal flux estimated at the Sun (with flare ribbons and post-eruption arcades, respectivelly) and at 1 au  (within MCs) because the studied fluxes span a factor about 30, so that even an error of a factor 2 on individual cases has a relative moderate impact on the correlation.   

   %{\S}{\bf --- Saturation of $\fz$} \\
The function $\fz (r)$ is peaked at $r=1$ and it has a behavior similar than $\ba (r)$ with the difference that the extra factor $r$ provides lower values near the origin.   This implies the transformation of the distributions of $\rsup$ and $\rinf$ (\figS{4_histo_r}a,c) to ones of \figS{histo_fa_fz}b,e.  Both show the clustering below $\fz=1$ and the extended tail towards low values.  Then, the distributions of $\fzsup$ and $\fzinf$ are similar to those of $\basup$ and $\bainf$ (\figS{histo_ba_bz}a,e), respectively, simply more extended on the low value side.  
  
   %{\S}{\bf --- More on $\fz$} \\
The effect of the function $\fz (r)$ on the MC results is further explicited in \fig{histo_fa_fz}f when compared to \fig{2_r}e.  A large clustering of both $\fzsup$ and $\fzinf$ near unity is present for MCs with $\rsup$ and $\rinf$ spread around unity.  Next, for some $\rsup$ values larger than 1, $\fzsup < \fzinf$ (\fig{histo_fa_fz}f). This is because $\fz (r)$ is not monotonically increasing with $r$, and it has a maximum (\fig{histo_fa_fz}c). 
Finally the low $r$ values, mostly $\rinf$ values, imply a dispersion in a long tail of low $\fz$ values.  In summary, all the $r$ values around unity contribute to the large peak of the $\fzsup$ and $\fzinf$ distributions, while $r$ values significantly away unity, both below and above, contribute to the long and weak tail of $\fz$ distribution (\fig{histo_fa_fz}b,e). 

   %{\S}{\bf --- More on $\fz$} \\
 In conclusion, because both $\fzsup$ and $\fzinf$ distributions are mostly clustered near unity, the axial flux is well estimated for most MCs, \ie\ nearly independently of the precise location of the MC boundaries  (within the limits of the Lundquist approximation of the FR, \fig{histo_fa_fz}b,e).  This contrasts with the distributions of azimuthal flux which are broad (\fig{histo_fa_fz}a,d).

  \begin{figure}    %%%%%%%%%%%%%%%%%% FIGURE 9   twist 
  \centering
  \includegraphics[width=\textwidth,clip=]{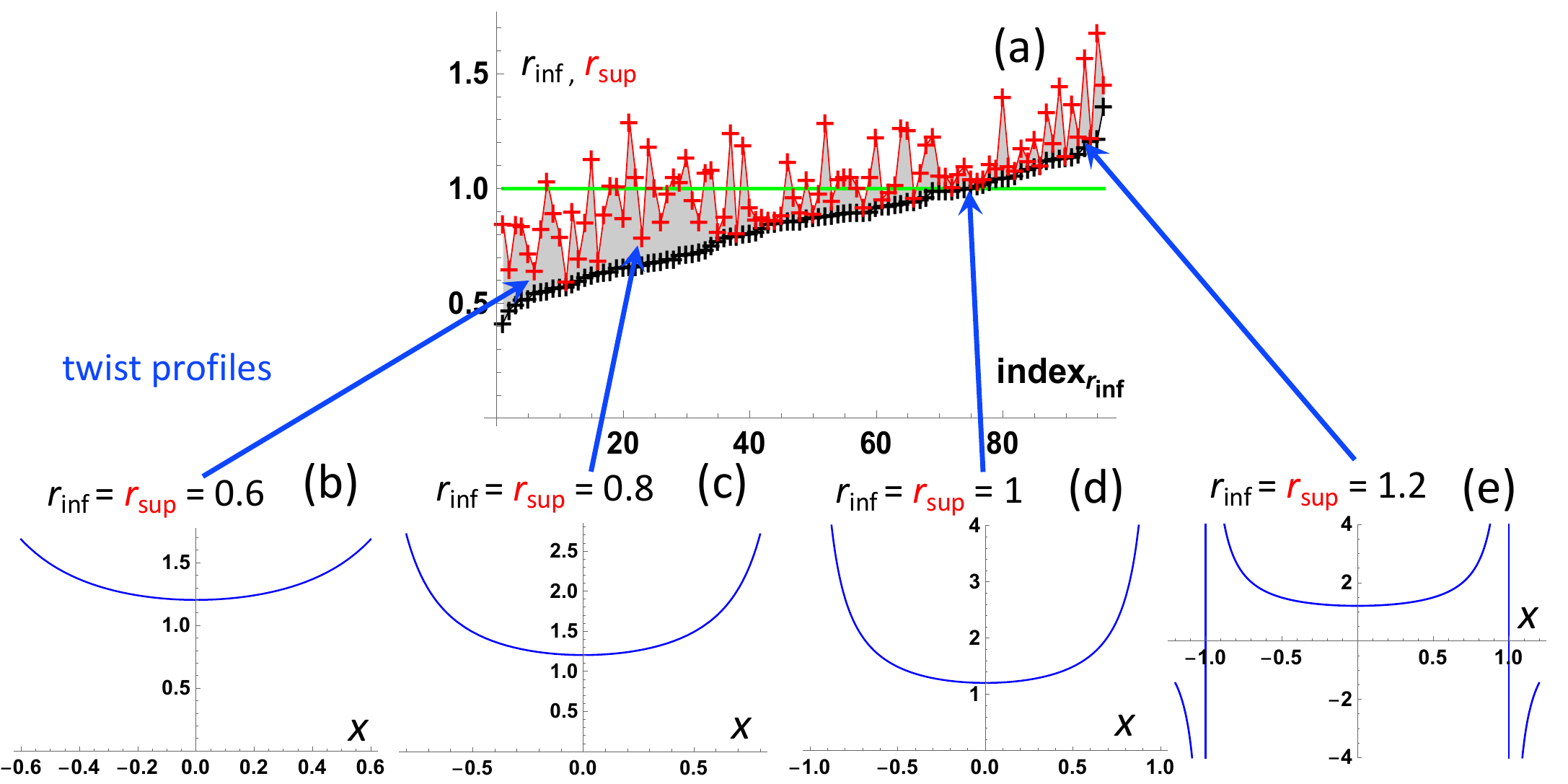} 
  \caption{ (b-e) Example of twist profiles obtained with the Lundquist's model set in the context of \fig{2_r}f shown at the top (a).  The examples are selected to be simple with symmetric boundaries ($\rinf=\rsup$) and with a trajectory crossing the FR axis ($p=0$).  They are still representative of the variety of the observed cases as indicated by the blue arrows. 
          }
  \label{fig_twist_profiles}
  \end{figure}
  
\subsection{Magnetic Twist} %%%%%%%%%%%%%%%%%%%%%%%%%%%%%
  \label{sect_Implications_Twist} 
  
   %{\S}{\bf --- Implication for twist profiles} \\
The broad distributions of $\rinf$ and $\rsup$, \fig{4_histo_r}, also imply a variety of twist profiles as shown in \fig{twist_profiles}.   We selected simple cases with $p=0$ and $\rinf=\rsup$, which are still representative of the variety of cases. 
Cases with $|p|>0$ have flatter twist profiles in function of $X$ (as the FR core is not crossed), while cases with $\rinf$ lower than $\rsup$ only corresponds to profiles more truncated on the $\rinf$ side.  The twist profile shown is located in the $\rinf,\rsup$ plot \versus\ the index of $\rinf$ (same as \fig{2_r}f).  The four selected cases have nearby results of MCs (crosses).  

   %{\S}{\bf --- Usual case} \\
The case $\rinf=\rsup =1$ is the one described by Lepping in most of his papers since the analyzed FR model extends up to $\Ro$ (\eq{Ro}).   Moreover, this is the case explicitly used in the fit of data by authors imposing $\BzFR =0$ at the MC boundaries \citep[\eg ][]{Lynch03, Lynch05, Vandas06, Wang15, Good19}.  This implies that the twist diverge to infinity at the MC boundaries (\fig{twist_profiles}d).  
       
   %{\S}{\bf --- Broad range of twist profiles} \\
As $\rinf=\rsup$ is lower, the twist profile is flatter (\figS{twist_profiles}b,c) which is expected since only a more central part of the FR core is used.  At the opposite, for $\rinf=\rsup>1$, the twist profile has an infinite branch at $r=1$, which corresponds to the reversal of $\BzFR$ (\fig{twist_profiles}e).   

   %{\S}{\bf --- Conclude on twist profiles} \\
These four examples of twist profiles, with observed cases all around (arrow heads), scan the typical twist profiles deduced from \insitu\ observations.  We conclude that the twist profile deduced from fitting the Lundquist solution to \insitu\ data has a broad range of twist profiles when limited to the time interval of observed MCs.   

   %{\S}{\bf --- comparison to Gold and Hoyle} \\
Another model with cylindrical symmetry is a non-linear FFF with a uniform twist \citep{Gold60}.  This model was fitted successfully to large sets of MCs \citep{Hu15,Wang16}.  They show that about half of the studied MCs can be approximately fitted with the Gold and Hoyle model.  Is this model better to represent the FR within MCs?  
In fact, fitting the Lundquist's model to \insitu\ data with the Lepping procedure includes the possibility of a uniform twist if the data have a uniform twist.  The fit will simply return  large values of $\Ro$, much larger than $\Rsup$. Then, using the Lundquist's model up to $\Ro$ would provide a large extrapolation of the magnetic field outside the observed MCs.  However, within the analysed MC, the Lundquist and the Gold and Hoyle models would provide very nearby fits.  Since both models have the same number of free parameters, we conclude that the Lundquist model should be preferred to fit \insitu\ data with the Lepping's procedure.
   
   %{\S}{\bf --- Twist in the FR core} \\
Assuming that the fit of the Lundquist solution is close enough to the data, \eg\ by limiting the sample to quality 1 and 2, Lepping's results imply a variety of twist profiles when limited to the MC time interval (\figS{twist_profiles}d-e).  Still, all these cases have in common a nearly uniform twist around the FR axis given by $\alpha/2$.   Then, Lepping's results
disagree with the ones of \citet{Wang16,Wang18} who claim that a higher twist exist in the core (compared with the surrounding).  Moreover, since $\Ro =c/\alpha$ (with $c\approx 2.4$), $\Ro$ is intrinsically linked to the core twist and not to the radius of the FR crossed by the spacecraft (in general $\Rinf, \Rsup \neq \Ro$).
 
  \begin{figure}    %%%%%%%%%%%%%%%%%% FIGURE 10     H 
  \centering
  \includegraphics[width=\textwidth,clip=]{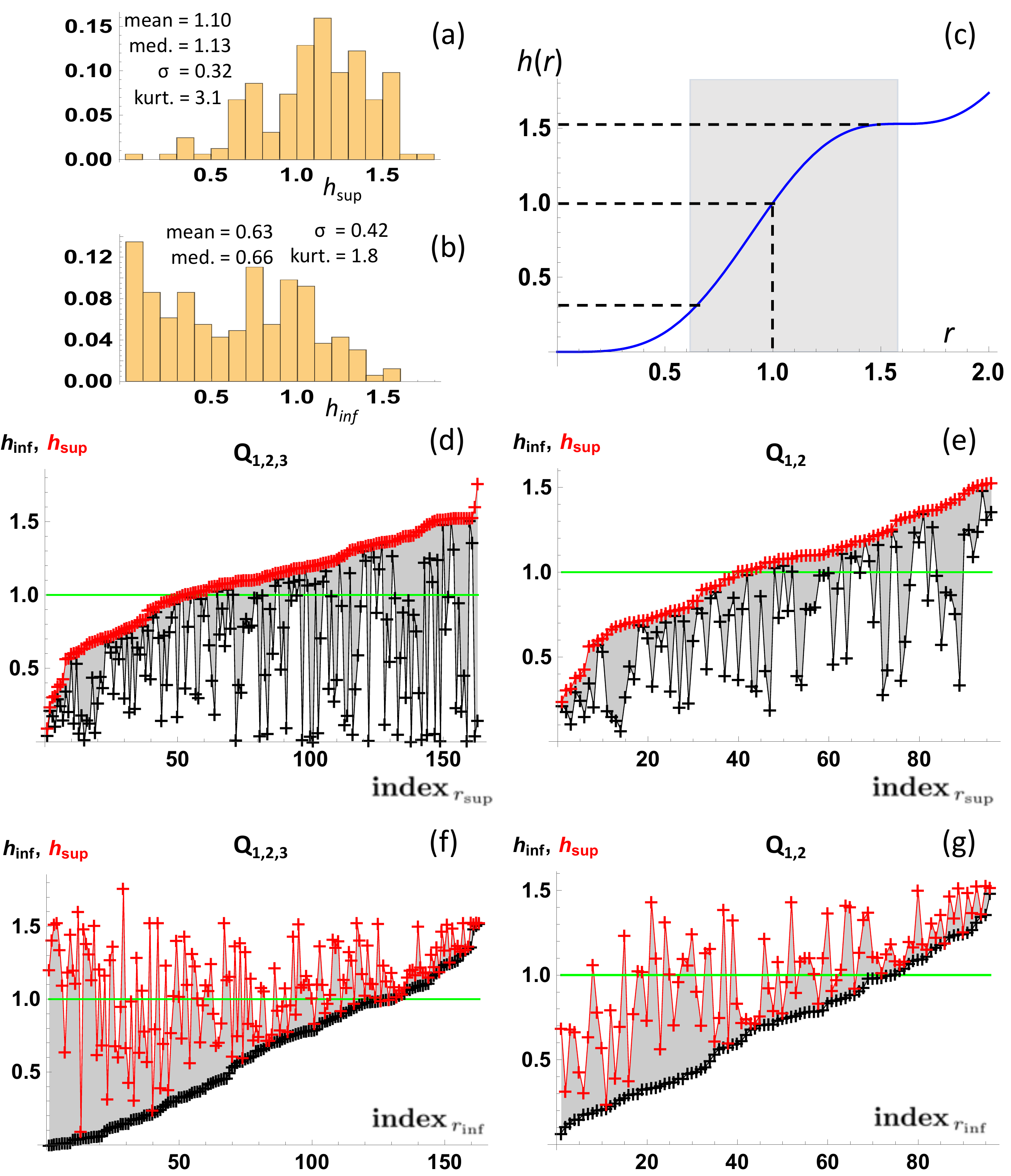} 
  \caption{ Normalized magnetic helicity $h$ of the fitted FRs with the boundary set at the MC boundaries, $\Rinf$ or $\Rsup$. The normalization is done with the helicity obtained with the FR boundary set at $\Ro$ (defined by $\BzFR=0$). 
    (a,b) Histograms of $\hsup$ and $\hinf$ for all MCs (quality 1,2,3).
    (c)   Dependance of the normalized magnetic helicity $h(r) = H(R)/H(\Ro)$ with $r=R/\Ro$.  The horizontal extension of the grey region is fixed to the mean $\pm 2 \sigma$ for $\rsup$ and Q$_{1,2,3}$ as in \fig{4_histo_r}.
    The four lower panels show $\hinf$ and $\hsup$ ordered in abscissa with $\rsup$ rank index (d,e), and with $\rinf$ rank index (f,g).  (d,f) show all MCs, and (e,g) only MCs of quality 1 and 2.
           }
  \label{fig_H}
  \end{figure}

\subsection{Magnetic Helicity} %%%%%%%%%%%%%%%%%%%%%%%%%%%%%
  \label{sect_Implications_Helicity}      

   %{\S}{\bf --- Magnetic helicity} \\
 The magnetic helicity $H$ quantifies how all the elementary magnetic flux tubes are winded around each other in a defined volume.
 The helicity ($H$) of a straight flux rope is the summation of the linked fluxes $\Fz$ and $\rmd \Fa$ with the fluxes defined in \eq{Fa_Fz} \citep{Berger03b,Dasso05d}:
  \begin{equation}  \label{eq_H}
  H(\rho) = 2 \, \int_{0}^{\rho} \Fz (\rho') \frac{\rmd \Fa (\rho') }{\rmd \rho'} \rmd \rho'   \,,
  \end{equation}
where $\Fa$ and $\Fz$ are the azimuthal and axial magnetic flux, respectively. 
Next, we define $h(r)$ by normalizing $H(\rho)$ with $H(\Ro)$ and $r=\rho/\Ro$.
After an analytical integration, with $\fz $ and $\fa$ defined by \eq{fa_fz}, $h(r)$ writes
  \begin{equation}  \label{eq_hr}
  h(r) = 2 \, \int_{0}^{r} \fz (r') \frac{\rmd \fa (r') }{\rmd r'} \rmd r'
       =  r^2 \, [ J_1^2(\alpha r) - J_0(\alpha r) \, J_2(\alpha r) ] \,\,  J_1^{-2} (\alpha) \,.
  \end{equation}

   %{\S}{\bf --- Interpretation of $h(r)$} \\
$h(r)$ is weak in the FR core (\fig{H}c) since there is a weak azimuthal flux (associated to a weak $\Ba$) wrapping around a weak axial flux (as the surface involved scales as $r^2$, so that even with large values of $\Bz$ in the core, $\fz $ stays small, \fig{histo_fa_fz}c).  Around $r=1$, $h(r)$ has a linear behavior with $r$ as this region is associated to a maximum of $\fz(r)$ and a linear increase of $\fa(r)$. $h(r)$ reaches a first local maximum where both $\rmd \fa /\rmd r \propto \ba(r)$ and $\fz(r) \propto r \,\ba(r)$ vanish.  Then, both contributions in the integral of \eq{H} change of sign together, and more generally the contribution of the FR layers are always of the same sign (those of $\alpha$). This is in contrast with the twist profile that reverse its sign at $\Ro$, or $r=1$ (\fig{twist_profiles}).

   %{\S}{\bf --- Implication for $\hsup$} \\
$\rsup$ values are mainly located in the grey band of \fig{4_histo_r} (defined as mean $\pm 2 \sigma$ of $\rsup$), and reported in \fig{H}c.  With the exception of the most right part of the interval, this corresponds to the range of the linear variation of $h(r)$, then the values of $\hsup$ are mostly a rescaled version of $\rsup$ values, with simply a broader distribution (comparing \fig{H} to \figs{2_r}{4_histo_r}).  The larger difference is for the large $\rsup$ values where $h(r)$ is flatter, leading to a saturation effect. 
This implies that the steeper slope, present for the larger indexes of $\rsup$ in \figS{2_r}c,d, disappears in \figS{H}d,e. Then, the right tail of $\hsup$ histogram is almost not present (\fig{H}a).  In contrast, the left tail of $\hsup$ histogram is extended compare to the one of $\rsup$.  We describe this effect below since it is stronger for $\hinf$.

   %{\S}{\bf --- Implication for $\hinf$ } \\
  Since $\rinf$ extends to lower values than $\rsup$, the non-linearity of $h(r)$, for small $r$ values, implies very low values for $\hinf$ for the smaller $\rinf$ values (\figS{H}f,g).  The helicity computed with $\rinf$ has a broad range of values compare to the one computed with an ideal FR ($\BzFR=0$ at the FR boundary). Indeed, the histogram of $\hinf$ peaks around 0 (\fig{H}b)
and is very different from the histogram of $\hsup$ which rather peaks just above 1 (\fig{H}a).  Since $\rinf$ defined approximately the FR which remains at 1 au while $\rsup$ is an estimation of the FR close to the Sun, we conclude that the erosion removes an important amount of magnetic helicity during the travel from the Sun to 1 au. The same conclusion is reached  if MCs are restricted to quality 1 and 2 (\figS{H}f,g are similar).

  \begin{figure}    %%%%%%%%%%%%%%%%%% FIGURE 11   p 
  \centering
  \includegraphics[width=\textwidth,clip=]{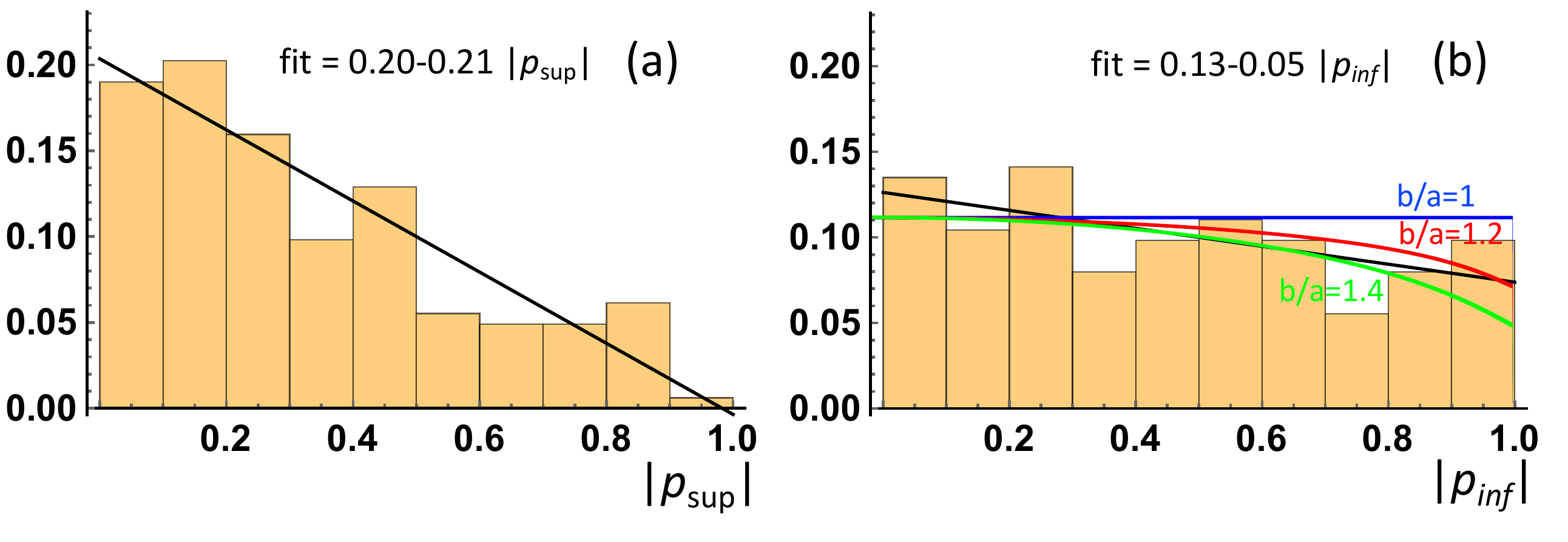} 
  \caption{ Distributions of the impact parameters (a) $|\psup| = |p| /\rsup$ and 
  (b) $|\pinf| = |p| /\rinf$.   The distributions are least square fitted with a linear function (black line). The color curves are the distributions computed with \citet{Vandas03} model and an aspect ratio, $b/a$, of the major to minor axis of the cross-section as reported.
           }
  \label{fig_p}
  \end{figure}

\subsection{Impact Parameter} %%%%%%%%%%%%%%%%%%%%%%%%%%%%%
  \label{sect_Implications_p}      

   %{\S}{\bf --- Distribution of $|p|$. Previous studies.} \\
The probability distribution of the absolute value of the impact parameter $|p|=|CA|/100$ was found to decrease strongly with $|p|$ \citep{Lepping10}.
This contrasts with the expectation of an approximately flat distribution as the consequence of spacecraft and MC trajectories having no relationship (so all $|p|$ values are expected to be equiprobable).   Using various FR models, \citet{Demoulin13} demonstrated that a non-detection of MCs due to a lower magnetic field strength and lower field rotation for larger $|p|$ crossings could not explain the observed distribution of $|p|$.  Rather, the observed $|p|$ distribution of MCs could be explained by a FR cross section elongated in the direction transverse to the spacecraft trajectory, on average by a factor 2 to 3 depending on the magnetic twist profile. 

   %{\S}{\bf --- Need to redefine $p$} \\
The study of \citet{Demoulin13} is based on Lepping's results, which report the fitted FR with radius $\Ro$ (\eq{Ro}).  But using $\Ro$ implies extrapolating the FR model outside of the observed MC for the majority of cases (cases with $\rinf <1$ in \fig{4_histo_r}).   For a small number of MCs, $\rinf >1$, which allows the possibility of a spacecraft trajectory passing at a distance in between $\Ro$ and $\Rinf$, so having $|p|>1$. A few cases are present in Lepping's table.  Depending on the purpose, we rather propose that $\Rsup$ or $\Rinf$ should be used to define the impact parameter, so we define:  $\psup = p \Ro/\Rsup$ and $\pinf = p \Ro/\Rinf$.

   %{\S}{\bf --- Distribution of $\psup$} \\
Since $\rsup$ is fluctuating around 1 (\figs{2_r}{4_histo_r}), the distribution of $|\psup|$, \fig{p}a, is similar to the one of $|p|$ \citep[Figure 1a in][]{Demoulin13}.  The least square fit to a linear function provides the global tendency which also is similar (with slope only steeper by 10\% for $|\psup|$ than for $|p|$).  Comparable results also are obtained by restricting the sample to quality 1 and 2. Then, within the precision of the method, the results of \citet{Demoulin13} apply to the FRs defined by $\Rsup$.  

   %{\S}{\bf --- Distribution of $\pinf$} \\
The impact parameter also could be defined with $\Rinf$.   This is a safer way as it represents the remaining FR when the spacecraft crossed the MC. Since $\rinf$ is typically lower than 1 and broadly distributed (\figs{2_r}{4_histo_r}) the distribution of $|\pinf|$ is significantly different (\fig{p}b) with a much flatter distribution than the one of $|p|$.  Following the analysis of \citet{Demoulin13}, with the linear FFF model of \citet{Vandas03} which generalise the Lundquist's model to elliptical cross section, we conclude that the remaining FRs at 1 au are typically close to a circular cross section, with typically an aspect ratio around $1.3$ (\fig{p}b). 

   %{\S}{\bf --- Coherence with previous studies} \\
  Using combined observations of several spacecraft, some previous analyses have shown that the core of MCs is significantly more circular than their oblate outer part \citep{Liu08b,Kilpua09,Mostl09}.   The above result with $|\pinf|$ distribution shows that is likely typical at 1 au even if some MCs have a flatter cross section \citep[\eg ][]{Vandas05}.
     
%%%%%%%%%%%%%%%%%%%%%%%%%%%%%%%%%%%%%%%%%%%%%%%%%%%%%%%%%%%%%%%%%%%%%%%%%%%%%%%%%%
\section{Conclusion} %%%%%%%%%%%%%%%%%%%%%%%%%%%%%%%%%%%%%%%%
      \label{sect_Conclusion} 
      
  %{\S}{\bf --- Summarize the global aim} \\
The \insitu\ observations provide a 1D cut across magnetic structures such as MCs.
The global understanding of these data is not unique but the accumulation of various studies point to the detection of FRs.  This is supported by the relatively good fit of FR models and by our present understanding of the physics of their solar origin. 
Having a reliable method to derive the FR local orientation is important as the associated errors affect all other global quantities such as size, magnetic fluxes and helicity.   Such parameters are important to link quantitatively \insitu\ observations to their solar sources.  More precise physical parameters are also important to better understand the physics of the transport from the Sun, in particular the importance of magnetic erosion. 

  %{\S}{\bf --- Strength of Lepping procedure} \\
In this paper, we focus on the method of Lepping \citep{Lepping90} both because it is the most widely applied to \insitu\ data and because it was a seminal method used as a base to develop other fitting methods.
The method of Lepping is based on the Lundquist model.  It involves a minimum number of free parameters needed for any FR model fitted to the data if no constraint is imposed (\eg\ not imposing the vanishing of the axial component, $\BzFR$, at the MC boundaries).
Said differently, the above fit has no intrinsic parameter provided by the Lundquist's model since $\alpha$ (or equivalently $\Ro$) and $\Bo$ in \eq{Lundquist} are just rescaling parameters in spatial size and field strength, respectively. Other parameters are the helicity sign and the geometrical parameters of the encounter (axis orientation angles, time and closest approach distance from the axis). 
  The method of Lepping can be transposed to other models, \eg\ as it was done with the Gold and Hoyle model, which introduces the same number of free parameters than the Lundquist's model.  A more elaborated model could be fitted by the same method, with the inclusion of extra parameters intrinsic to the model.
 We conclude that the fitting method of Lepping is general and well justified.
   
  %{\S}{\bf --- Problem: $\Ro$ was emphasized! No justification!} \\
One key parameter of the fit is the spatial scaling parameter, $\Ro$, defined by the radius where $\BzFR =0$ (vanishing axial component).
   First, we notice that $\Ro$ only scales the model to the data, and it could have been defined at any other value of $\BzFR$ without affecting the fitted model.
This different definition is indeed needed for Gold and Hoyle's model since $\BzFR$ is not vanishing at any finite radius.
   Second, we show that using $\Ro$ as the FR boundary defines a FR which could be smaller or larger than the studied MC with different extensions of the in- and outbound regions.   Moreover, this selection of $\Ro$ is in general not justified by \insitu\ data, nor by the physics of the solar launch.  At most, it is justified in the ideal case where a FR is launched from a coronal magnetic configuration, invariant by translation and embedded in a potential field arcade which fully reconnects behind the erupting FR.   Still, solar eruptive configurations are more divers than this, so a distribution of the axial field component, $\BzFR$, is rather to be expected.
This distribution is further broaden by the variety of magnetic erosion occurring during the travel from the Sun to 1 au.   

  %{\S}{\bf --- $\BzFR=0$ from other authors} \\
While $\Ro$ is one of the fitted parameter in Lepping's method, we emphasise that $\BzFR=0$ is not imposed at the MC boundaries in contrast with other methods which explicitly set $\BzFR=0$ at the MC boundaries \citep[\eg][]{Lynch03, Lynch05, Vandas06, Wang15, Good19}.  
\citet{Nishimura19} show that imposing or not $\BzFR =0$ at the MC boundaries could have an important effect on the computed axis orientation. 
This is likely the main origin of the difference in FR orientations found by \citet{Lynch05} and \citet{Lepping10} for the same MCs as reported in Figure 2 of \citet{Janvier15}.  
    
  %{\S}{\bf --- Compute two FR radius} \\
In the case of Lepping's method the output parameters define the fitted model which, by construction, extends to the full 3D space, but which should only be considered within the MC time interval.   Then, from the fitted parameters we computed the FR radius at the two MC boundaries, defining $\Rin$ and $\Rout$.   These parameters replace $\Ro$ and $\to$ (\eq{Rin_Rout}) where $\to$ is the time of the closest approach of the spacecraft from the FR axis.   From Lepping's published results, $\to$ is only given indirectly with the absolute value of the asymmetry parameter asf.   This introduces an ambiguity on the location of $\to$, before or after the center of the MC time interval so that only the inferior, $\Rinf$, and superior, $\Rsup$, radius could be derived.  With these new parameters the matrix of results published by Lepping can be reloaded. This has several implications as follow.

  %{\S}{\bf --- MCs: not only a FR!} \\
At 1 au, $\Rinf$ and $\Rsup$ are typically different, showing that each MC is formed both by a FR and by an extra magnetic flux on one side. This extra flux could be accreted flux either at the Sun or during the travel to 1 au. However, this last case is unlikely since otherwise the physical properties of the accreted flux and plasma would be more solar wind like (as in the sheath in front of MCs).  This extra flux is rather expected to be mostly due to erosion on the other FR side, so it was part of the ejected FR earlier on, but not when observed at 1 au. This is in agreement with previous results on erosion \citep[][and references therein.]{Ruffenach15}.  In particular an example of each case, erosion from the front or from the rear, is shown in their Figure 2. 

  %{\S}{\bf --- Continuous distribution of FR radius} \\
The distributions of $\Rinf$ and $\Rsup$ provide observational constraints.
$\Rsup$ has a distribution spreading around $\Ro$ (where $\BzFR=0$, \fig{4_histo_r}) as expected from our knowledge of solar launch configurations and with the hypothesis of a weak erosion on one of the FR side.  By contrast, $\Rinf$ extends towards much lower values of $\Ro$, as expected with significant and variable erosion.
For both $\Rinf$ and $\Rsup$, there is no evidence of outsider cases but rather a quasi-continuous distribution of values (\fig{2_r}).   In particular, the so called annulus cases, with $\BzFR<0$ close to both MC boundaries, are not a separate class of events, but about a 20\% part of the MC distribution.  Also, $\BzFR<0$ at least at one boundary for about 60\% of MCs.
 
  %{\S}{\bf --- Consequences of different $\rinf$ and $\rsup$} \\
The distributions of $\Rinf$ and $\Rsup$ have implications on quantities depending on the FR boundaries.
   First, the axial field component $\bz$ (normalized to the central field strength $\Bo$) has a broad distribution both from data and from the Lundquist fit (especially for $\bzinf$: within [-0.5,1], \fig{histo_ba_bz}). 
   Second, the normalized azimuthal component near the MC boundaries is rather well defined ($\approx 0.5$) from the Lundquist fit while more broadly distributed from the data.  The magnetic fluxes derived from the Lundquist fit, normalized by their values obtained at $\Ro$, have mostly the opposite axial/azimuthal distributions: the axial flux is well defined while the azimuthal flux is broadly distributed (\fig{histo_fa_fz}).  Finally, magnetic helicity is the most affected quantity by the location of the FR boundaries, so by erosion during the travel from the Sun to 1 au (\fig{H}).

  %{\S}{\bf --- Implication for the twist} \\
The FR remaining at 1 au has a radius $\Rinf$.  
The twist profile is very significantly affected by the $\Rinf/\Ro$ value.  As it is lower, the twist profile is flatter (\fig{twist_profiles}).  Indeed, the Lepping's method includes the possibility to return a constant twist profile if the data include it.   Using the Lundquist model introduces the same number of free parameters as using the Gold and Hoyle model, so Lepping's method is superior as it allows to provide a constant twist profile without forcing it.  In fact the strength of Lepping's method is that, while the minimum number of free parameters is used, it allows a relatively broad range of possible twist profiles.  This allows to have at least an approximate twist profile which range from flat to increasing away the FR axis, up to infinity.   Further it allows the reversal of the axial field component.   All these configurations are expected from the solar source properties and with erosion, so Lepping's method is robust because it incorporates approximately, with a minimum number of parameters, the consequences of the main physical ingredients of the FR build up and transport.  
   
  %{\S}{\bf --- Consequences for impact parameter} \\
The closest approach distance CA is expressed as a percentage of $\Ro$ in Lepping's results, or as a fraction of $\Ro$ with the impact parameter $p$.  Since $\Rsup$ has a distribution centered around $\Ro$, $p$ recomputed as a fraction of $\Rsup$ has a similar distribution than the original one.  However the remaining FR at 1 au has a radius $\Rinf$.  $p$ recomputed as a fraction of $\Rinf$ has a nearly uniform distribution as expected for randomly distributed encounters of FR ropes with circular cross-sections. 
More precisely, an elliptical linear FFF model with an aspect ratio $\approx 1.3$,  is sufficient to interpret the distribution of $p$ computed with $\Rinf$, then the cross-sections are, in average, close to circular sections.

  %{\S}{\bf --- Getting deeper with \insitu\ observations closer from the Sun } \\
Finally, we conclude that the Lepping's method includes all the key elements for fitting a FR model to \insitu\ data of MCs.  It is an economic fit as it includes the minimum number of free parameters without imposing unphysical conditions.
The selection of the Lundquist model is particularly judicious as it includes a variety of twist profiles which are globally expected from solar physics.   
The key point is to understand the meaning of the fitted model, in particular its range of validity, which is set by the MC boundaries defined on the data. This implies to redefine some of the output parameters in order to keep a model representing the crossed MC and not a fictitious associated FR (which could be smaller or larger than the studied MC).
With this approach, we believe that Lepping's method will still be very useful to analyze data of other missions such as ESA’s Solar Orbiter and NASA’s Parker Solar Probe, and to investigate the evolution of MCs in the interplanetary medium.

%%%%%%%%%%%%%%%%%%%%%%%%%%%%%%%%%%%%%%%%%%%%%%%%%%%%%%%%%%%%%%%%%%%%%%%%%%%%%%%%%%
\begin{acks}
We thank the referee for his/her comments which broaden the potential audience of the paper.
%% Grants
S.D. acknowledges partial support from the Argentinian grants UBACyT (UBA), and PIP-CONICET-11220130100439CO.
This work was partially supported by a one-month invitation of P.D. to the Instituto de Astronom\'ia y F\'isica del Espacio,  
and by a one-month invitation of S.D. to the Observatoire de Paris.
This work was supported by the Programme National PNST of CNRS/INSU co-funded by CNES and CEA. 
%% Conicet
S.D. is member of the Carrera del Investigador Cien\-t\'\i fi\-co, CONICET.
\end{acks}\\

\noindent {\footnotesize \textbf{Conflict of Interest:} The authors declare that they have no conflict of interest.}

%%%%%%%%%%%%%%%%%%%%%%%%%%%%%%%%%%%%%%%%%%%%%%%%%%%%%%%%%%%%%%%%%%%%%%%%%%%
%\appendix   

     % format of references provided by the journal (.bst)
\bibliographystyle{spr-mp-sola}
     % name your Bibtex file containing your references (.bib)
\bibliography{mc}  

     % Checking: look if the file containing the ``\bibitem'' exits
     %           so check if the .bbl file exist (bibTeX compilation)
\IfFileExists{\jobname.bbl}{} {\typeout{}
\typeout{****************************************************}
\typeout{****************************************************}
\typeout{** Please run "bibtex \jobname" to obtain} \typeout{**
the bibliography and then re-run LaTeX} \typeout{** twice to fix
the references !}
\typeout{****************************************************}
\typeout{****************************************************}
\typeout{}}

\end{article} 

\end{document}